\documentclass[a4paper,fleqn,usenatbib,useAMS]{mnras}

\usepackage{graphicx}	
\usepackage{amsmath}	
\usepackage{amssymb}	
\usepackage{multicol}   
\usepackage{bm}		
\usepackage{pdflscape}	
\usepackage[export]{adjustbox}
\usepackage{cancel}
\newcommand{\ct}{\citealt}
\renewcommand{\thesubsubsection}{\thesubsection.\alph{subsubsection}}

\defcitealias{Zhao+2020}{Paper I}
\newcommand{\PaperI}{\citetalias{Zhao+2020}}

\usepackage[T1]{fontenc}
\usepackage{ae,aecompl}

\usepackage{newtxtext,newtxmath}

\title[Non-ideal MHD \& Disc Formation]{The Interplay between Ambipolar Diffusion and Hall Effect on Magnetic Field Decoupling and Protostellar Disc Formation}

\author[B. Zhao et al.]{Bo Zhao$^{1}$\thanks{Contact e-mail: \href{mailto:bo.zhao@mpe.mpg.de}{bo.zhao@mpe.mpg.de}}\thanks{Present address: Giessenbachstr. 1, D-85748, Garching, Germany},
Paola Caselli$^{1}$,
Zhi-Yun Li$^{2}$,
Ruben Krasnopolsky$^{3}$,
Hsien Shang$^{3}$,
Ka Ho Lam$^{2}$\\
\\
$^{1}$Max-Planck-Institut f\"{u}r extraterrestrische Physik (MPE), Garching, Germany, 85748\\
$^{2}$University of Virginia, Astronomy Department, Charlottesville, USA, 22904\\
$^{3}$Academia Sinica Institute of Astronomy and Astrophysics, 10167, Taipei, Taiwan\\}

\pubyear{2019}

\begin{document}
\label{firstpage}
\pagerange{\pageref{firstpage}--\pageref{lastpage}}
\maketitle

\begin{abstract}
{Non-ideal MHD effects have been shown recently as a robust mechanism of averting 
the magnetic braking ``catastrophe'' and promoting protostellar disc formation. 
However, the magnetic diffusivities that determine the efficiency of non-ideal MHD 
effects are highly sensitive to microphysics. We carry out non-ideal MHD 
simulations to explore the role of microphysics on disc formation and the interplay 
between ambipolar diffusion (AD) and Hall effect during the protostellar collapse. 
We find that removing the smallest grain population ($\lesssim$10~nm) from the 
standard MRN size distribution is sufficient for enabling disc formation. 
Further varying the grain sizes can result in either a Hall-dominated or an 
AD-dominated collapse; both form discs of tens of AU in size regardless of the magnetic 
field polarity. The direction of disc rotation is bimodal in the Hall dominated collapse 
but unimodal in the AD-dominated collapse. We also find that AD and Hall effect 
can operate either with or against each other in both radial and azimuthal directions, 
yet the combined effect of AD and Hall is to move the magnetic field radially outward 
relative to the infalling envelope matter. In addition, microphysics and magnetic field 
polarity can leave profound imprints both on observables (e.g., outflow morphology, disc to 
stellar mass ratio) and on the magnetic field characteristics of protoplanetary discs. 
Including Hall effect relaxes the requirements on microphysics for disc formation, 
so that prestellar cores with cosmic-ray ionization rate of 
$\lesssim$2--3$\times10^{-16}$~s$^{-1}$ can still form small discs of $\lesssim$10~AU 
radius. We conclude that disc formation should be relatively common for typical 
prestellar core conditions, and that microphysics in the protostellar envelope is 
essential to not only disc formation, but also protoplanetary disc evolution.}
\end{abstract}




\section{Introduction}
\label{Chap.Intro}

Protostellar disc formation is a critical step between the collapse of dense molecular 
cores and the formation of stars and planets. How rotationally supported discs 
(RSDs hereafter) are formed from magnetized dense cores remains 
an unsettled question in existing literature. The main debate is on how 
to avert the ``catastrophic'' magnetic braking that transports away angular 
momentum from the circumstellar region and hence suppresses disc formation 
\citep{Allen+2003,MellonLi2008,HennebelleFromang2008}. Potential solutions that 
have been proposed in recent years include: misalignment between the initial magnetic 
field and rotation axis \citep{Joos+2012,Li+2013}, initial turbulence 
\citep{Santos-Lima+2012,Seifried+2013,Li+2014}, and non-ideal MHD effects 
\citep{Masson+2016,Tomida+2015,Tsukamoto+2015a,Tsukamoto+2015b,Zhao+2016,Wurster+2016,Zhao+2018a}. 
For the former two candidates, however, either a large misalignment angle \citep{Li+2013} 
or a sonic turbulence Mach number \citep{Li+2014} are needed initially for RSDs 
to form and survive from dense cores magnetized to a realistic level 
\citep[dimensionless mass-to-flux ratio of a few;][]{TrolandCrutcher2008}. 
The large misalignment angle or turbulence level is unlikely to be the typical 
condition for dense cores. Kinematic studies of prestellar cores have shown that the 
level of turbulence in dense cores is generally sub-sonic 
\citep{FullerMyers1992,Caselli+2002,KetoCaselli2008}. 
As dense cores are only slightly ionized \citep{Caselli+1998,BerginTafalla2007}, 
the flux-freezing conditions in the ideal MHD limit should no longer hold 
during the protostellar collapse, and non-ideal MHD effects should naturally 
operate in dense cores. 

The efficiency of the non-ideal MHD effects, especially ambipolar diffusion (AD) 
and Hall effect, in regulating the protostellar collapse and disc formation, depends 
heavily on the ionization fraction and microphysics in dense cores. Early non-ideal 
MHD studies on disc formation have adopted relatively low magnetic diffusivities, 
which lead to the general conclusion that disc formation remain suppressed 
\citep[e.g.,][]{Machida+2007,MellonLi2009,Li+2011}. In particular, AD can instead 
enhance the magnetic field strength and hence the magnetic braking in the inner 
envelope by driving a hydrodynamic C-shock that moves radially outward into 
the infalling flow 
\citep[so-called ``AD-shock'';][]{LiMcKee1996,KrasnopolskyKonigl2002,Li+2011}. 
The formation of AD-shock owes to a negligible decoupling of magnetic fields 
in the bulk envelope and an abrupt decoupling in the stellar vicinity, which 
could be mostly avoided by a larger ambipolar diffusivity in the envelope 
\citep{KrasnopolskyKonigl2002,Zhao+2018a}. Nevertheless, the general consensus 
is that magnetic diffusivities in the collapsing envelope have to be enhanced by 
$\sim$1--2 orders of magnitude than the values adopted in these early studies, 
so as to enable the formation of tens-of-AU RSDs \citep{Shu+2006,Krasnopolsky+2010}. 

The main microphysical properties that control the magnetic diffusivities 
are cosmic-ray (CR) ionization rate and grain size distribution 
\citep[e.g.,][]{Padovani+2014,Zhao+2016}. The CR ionization rate affects the 
overall magnitude of magnetic diffusivities \citep{UmebayashiNakano1990} while 
a large population of very small grains (VSGs: $\sim$1~nm to few 10~nm) can 
dominate the fluid conductivity \citep{Zhao+2016,Dzyurkevich+2017}. 
\citet{Dapp+2012} combine AD and Ohmic dissipation into an effective resistivity 
and explore different grain sizes for the resistivity computation. Their result 
reveals a strong dependence of the combined resistivity on grain sizes, 
especially at envelope densities ($\lesssim$10$^{10}$~cm$^{-3}$). 
However, the somewhat high CR ionization rate ($5 \times 10^{-17}$~s$^{-1}$) 
and slow initial rotation adopted in their study likely prevent the formation of 
sizable discs. Later, detailed investigations of the impact of grain size 
distribution on magnetic diffusivities have been carried out 
\citep{Padovani+2014,Zhao+2016,Dzyurkevich+2017,Koga+2019}, confirming the trend 
found in \citet{Dapp+2012} that slightly increasing the average grain size can 
enhance the ambipolar diffusivity by $\sim$1--2 orders of magnitude in comparison 
to the standard MRN \citep[Mathis-Rumpl-Nordsieck;][a -3.5 power law with size ranging from $a_{\rm min}$$\sim$0.005~$\mu$m to $a_{\rm max}$$\sim$0.25~$\mu$m]{Mathis+1977} size distribution. 
As shown analytically and numerically, the smallest grains are rapidly 
depleted in cold dense environment 
\citep{Ossenkopf1993,Hirashita2012,Kohler+2012,Silsbee+2020,Guillet+2020}, 
which is supported by the non-detection of spinning dust grain emission (produced 
by VSGs of $\lesssim$10~nm) in recent Galactic cold core surveys \citep{Tibbs+2016}. 
In fact, many recent non-ideal MHD simulations of disc formation have adopted 
grain size distributions free of VSGs for computing the magnetic diffusivities, 
for example, singly-sized 0.1~$\mu$m grains \citep{Tomida+2015}, or a ``modified'' 
MRN size distribution \citep{Marchand+2016,Masson+2016,Hennebelle+2020}, both of which 
would enhance the ambipolar diffusivity and promote disc formation.

Hall effect has recently been discussed extensively 
\citep{Krasnopolsky+2011,BraidingWardle2012a,BraidingWardle2012b,Tsukamoto+2015b,Wurster+2016,Marchand+2018,Zhao+2020} 
and claimed by several studies as the dominant mechanism for enabling a bimodal disc 
formation \citep[e.g.,][]{Tsukamoto+2017,WursterLi2018}, i.e, only when the angular velocity 
vector ($\bmath{\Omega}$) of the initial core is anti-aligned with the magnetic field 
($\bmath{B}$) that disc formation is possible. However, as pointed out by 
\citet[][hereafter \PaperI]{Zhao+2020}, the bimodality of disc formation by Hall 
effect does not continue into the main accretion phase; and in the absence of AD, Hall 
effect only allows the formation of $\sim$10--20~AU RSDs regardless of the sign of 
$\bmath{\Omega \cdot B}$. Particularly, in the aligned configuration 
($\bmath{\Omega \cdot B}$), both the disc and the inner envelope are counter-rotating 
with respect to the bulk core rotation. 
Moreover, the Hall diffusivity also benefits from removing the smallest 
$\lesssim$10~nm grains from the standard MRN size distribution, but reaches a maximum 
level at inner envelope densities when the minimum grain size is set to 
$\sim$0.03--0.04~$\mu$m \citep{Zhao+2018b,Koga+2019}. As shown in \PaperI, disc 
formation is strongly suppressed in models adopting the standard MRN size distribution, 
which is in agreement with the result of \citet{Li+2011}. In contrast, the grain size 
distribution in recent Hall studies are in general VSG-free 
\citep[e.g.,][]{Tsukamoto+2015b,Wurster+2016}, which naturally gives rise to 
efficient Hall effect. In particular, the grain size of 0.035~$\mu$m adopted by 
\citet{Tsukamoto+2015b} is very close to the average grain size needed for 
maximizing the Hall diffusivity, which causes Hall effect to dominate over 
AD in the collapsing envelope, as we will reveal in this study. 
Therefore, it is crucial to ensure the convergence of the microphysical properties 
and ionization chemistry before comparing the results of non-ideal MHD 
simulations of disc formation. 

The work of \citet{Dzyurkevich+2017}, \citet{Zhao+2018b}, and \citet{Koga+2019} 
have also discovered that ambipolar and Hall diffusivities behave differently 
when varying the grain sizes, 
and each diffusivity reaches its individual maximum level with a slightly different 
grain size ($a_{\rm min}$ at $\sim$0.1~$\mu$m for ambipolar but $\sim$0.03--0.04~$\mu$m 
for Hall). Such a difference can potentially allow AD and Hall effect to compensate 
each other as grain sizes change, when at least one of the two diffusivities is 
large enough. However, unlike AD that is diffusive along the bending direction of 
the magnetic field, Hall effect is a dispersive process that drifts the magnetic field 
lines along the orthogonal direction. Therefore, AD and Hall effect can interact in 
a non-trivial way depending on the relative importance of the two mechanisms. 
The impact of a varying ratio of Hall to ambipolar diffusivity on disc formation 
was first discussed by \citet{BraidingWardle2012a}, but only as a free parameter 
ranging from -0.5 to 0.2. In fact, the absolute value of Hall diffusivity can also 
become larger than the ambipolar diffusivity in the inner envelope when $a_{\rm min}$ 
is around 0.03--0.04~$\mu$m. In this paper, we will go beyond \PaperI~and elaborate 
on how microphysics changes the relative importance of AD and Hall effect, and on how 
the two effects interplay with each other during the protostellar envelope and 
disc formation. 

The rest of the paper is organized as follows. 
We demonstrate in Section~\ref{Chap.Non-ideal} the basic principles of non-ideal 
MHD effects in disc formation, and analyze in detail the ambipolar and Hall drift 
in the radial and azimuthal direction along the pseudo-disc; a generalized principle 
of the AD-Hall interplay is derived. Section~\ref{Chap.IC} describes the initial 
conditions of the simulation set, together with an overview of the results. 
In Section~\ref{Chap.SimulResult}, we start from the standard MRN size distribution, 
and demonstrate how the removal of the smallest nanometer-grain population can promote 
disc formation, and how further changes in microphysics can lead to either a
Hall-dominated collapse or an AD-dominated collapse. We show that disc formation 
is greatly promoted because of the persistent outward diffusion of magnetic fields 
in the radial direction. In Section \ref{Chap.Discuss}, we discuss the impact of 
microphysics on disc and outflow morphologies, and connect the process of 
disc formation to protoplanetary disc evolution. Finally, we summarize the results 
in Section \ref{Chap.Summary}.

\section{Non-ideal MHD Effects}
\label{Chap.Non-ideal}

The evolution of magnetic field $\bmath{B}$ in astrophysical fluids is governed 
by the magnetic induction equation, 
\begin{equation}
\label{Eq:induct}
\begin{split}
{\partial \bmath{B} \over \partial t} & = \nabla \times (\bmath{\varv} \times \bmath{B}) - \nabla \times \left\{\eta_{\rm O}\nabla \times \bmath{B} + \eta_{\rm H}(\nabla \times \bmath{B}) \times {\bmath{B} \over B}\right.\\
&\left.\hspace{74pt} +~\eta_{\rm AD}{\bmath{B} \over B} \times \left[(\nabla \times \bmath{B}) \times {\bmath{B} \over B}\right]\right\}\\
& = \nabla \times \left[(\bmath{\varv} + \bmath{\varv}_{\rm H} + \bmath{\varv}_{\rm AD}) \times \bmath{B} - \eta_{\rm O}\nabla \times \bmath{B}\right]~,
\end{split}
\end{equation}
where $\bmath{\varv}$ is the fluid velocity, $\eta_{\rm O}$, $\eta_{\rm H}$ 
and $\eta_{\rm AD}$ are the Ohmic, Hall, and ambipolar diffusivities, respectively; 
and $\bmath{\varv}_{\rm H}$ and $\bmath{\varv}_{\rm AD}$ denote the drift 
velocities of magnetic field lines induced by Hall effect and ambipolar diffusion, 
respectively, which are defined as, 
\begin{equation}
\label{Eq:v_H}
\bmath{\varv}_{\rm H} = -\eta_{\rm H} {\nabla \times \bmath{B} \over B} = -\eta_{\rm H} {4\pi \bmath{J} \over c B}~,
\end{equation}
\begin{equation}
\label{Eq:v_AD}
\bmath{\varv}_{\rm AD} = \eta_{\rm AD} {(\nabla \times \bmath{B}) \times \bmath{B} \over B^2} = \eta_{\rm AD} {4\pi \bmath{J} \times \bmath{B} \over c B^2}~,
\end{equation}
where $c$ is the light speed, and $\bmath{J}$ is the electric current. 
In what follows, we dissect the components of these drift velocities, as to 
better understand the role of non-ideal MHD effects in disc formation 
and evolution.

\subsection{Ambipolar \& Hall Drift}
\label{S.ADHallDrift}

Since ambipolar diffusivity $\eta_{\rm AD}$ is always positive 
\citep{WardleNg1999}, the direction of ambipolar drift in general points away from 
the bending direction of the magnetic field 
\citep[Fig.~\ref{Fig:sketch}; see also][]{Zhao+2018a}; namely, the ambipolar drift 
tends to relax the magnetic field bending, both 
radially and azimuthally, via magnetic tension force (Eq.~\ref{Eq:v_AD}). 
In the context of core collapse and disc formation, the bending of magnetic fields 
is the most severe along the pseudo-disc (usually the equatorial plane) or across 
the disc mid-plane. At such locations, the radial and azimuthal components of 
ambipolar drift velocity ($\varv_{{\rm AD},r}$ and $\varv_{{\rm AD},\phi}$) 
can be conveniently expressed in cylindrical coordinates, keeping only the 
leading terms, as,
\begin{equation}
\label{Eq:v_ADr}
\varv_{{\rm AD},r} \approx {\eta_{\rm AD} B_z \over B^2} {\partial B_r \over \partial z}~,
\end{equation}
and
\begin{equation}
\label{Eq:v_ADphi}
\varv_{{\rm AD},\phi} \approx {\eta_{\rm AD} B_z \over B^2} {\partial B_\phi \over \partial z}~,
\end{equation}
respectively; where $B_z$, $B_r$, and $B_\phi$ are the poloidal, radial, and azimuthal 
components of the magnetic field, respectively. It is clear that the ambipolar drift 
velocity is primarily determined by the magnetic field bending in the corresponding 
direction, i.e., $\varv_{{\rm AD},r} \propto {\partial B_r \over \partial z}$ 
and $\varv_{{\rm AD},\phi} \propto {\partial B_\phi \over \partial z}$. 
In particular, when the poloidal magnetic field lines are preferentially 
pinched inward, the radial ambipolar drift always points radially outward 
($B_z {\partial B_r \over \partial z} > 0$ in Eq.~\ref{Eq:v_ADr}).

In comparison, the Hall drift in a given direction is induced by the magnetic field 
bending in the orthogonal direction. As we have demonstrated in \PaperI, the radial and 
azimuthal components of the Hall drift velocity ($\varv_{{\rm H},r}$ and 
$\varv_{{\rm H},\phi}$) along the pseudo-disc or across the disc mid-plane 
can be estimated, keeping only the leading terms, as, 
\begin{equation}
\label{Eq:v_Hr}
\bmath{\varv}_{{\rm H},r} \approx {\eta_{\rm H} \over B} {\partial B_\phi \over \partial z}~,
\end{equation}
and
\begin{equation}
\label{Eq:v_Hphi}
\bmath{\varv}_{{\rm H},\phi} \approx -{\eta_{\rm H} \over B} {\partial B_r \over \partial z}~,
\end{equation}
respectively (see \PaperI~ for more detailed discussion). Basically, 
the Hall drift velocities are related to the magnetic field bending as 
$\varv_{{\rm H},r} \propto {\partial B_\phi \over \partial z}$ and 
$\varv_{{\rm H},\phi} \propto {\partial B_r \over \partial z}$. 
Note that Hall drift originates from the drift between positively and 
negatively charged species, thus $\eta_{\rm H}$ can be either positive 
(e.g., electrons drift relative to ions or positively-charged grains) 
or negative (e.g, ions drift relative to negatively-charged grains).

\subsection{Interplay between Non-Ideal MHD Effects}
\label{S.Interplay}

In the envelope of a collapsing core, both the ambipolar and Hall drift affect 
the evolution of magnetic fields, yet at slightly different scales. 
In general, AD can operate efficiently throughout most of the envelope 
\citep{Masson+2016,Zhao+2016,Zhao+2018a} while Hall effect only becomes 
efficient in the inner envelope (within a few 100~AU; \ct{Tsukamoto+2017}; \PaperI). 
The combined effect of AD and Hall can be represented by an effective drift velocity 
of the magnetic field lines with respect to the neutrals, 
\begin{equation}
\bmath{\varv}_{\rm d} = \bmath{\varv}_{\rm AD} + \bmath{\varv}_{\rm H}~.
\end{equation}
We define an effective velocity of the magnetic field lines $\bmath{\varv}_{\rm B,eff}$ as,
\begin{equation}
\bmath{\varv}_{\rm B,eff} = \bmath{\varv} + \bmath{\varv}_{\rm d}~,
\end{equation}
which can be substituted into Eq.~\ref{Eq:induct} to simplify the form of 
the induction equation.

\begin{figure*}
\includegraphics[width=\textwidth]{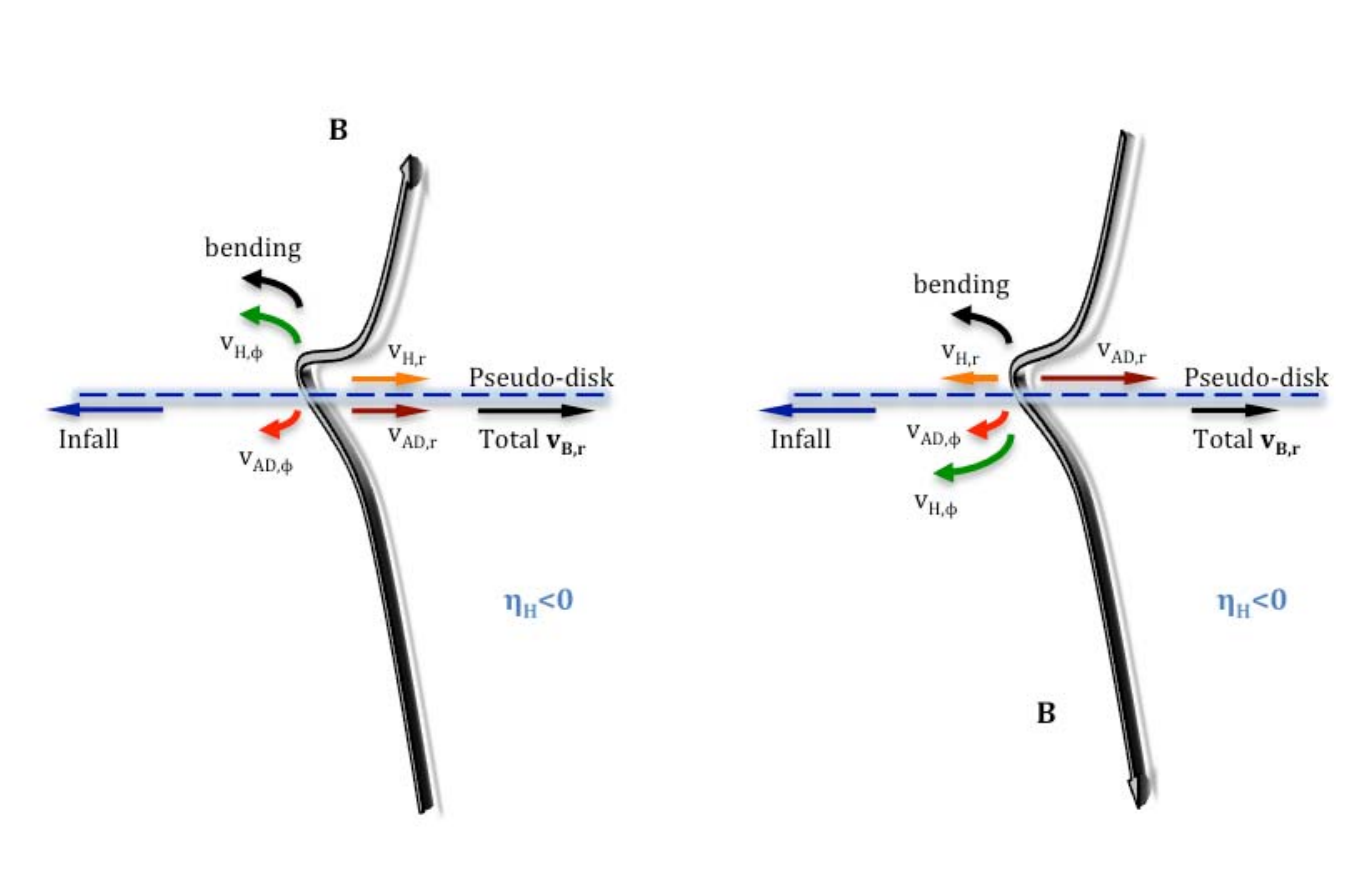}
\caption{Illustration of the interplay between Hall and AD effect in the collapsing 
envelope. In the radial direction, the combined effect of AD and Hall effect is to 
drift the magnetic fields radially outward. For the other two cases with the magnetic 
field being bended azimuthally towards the reader, the conclusions are similar. 
Note that the same principle can be applied to the protostellar disc, where magnetic 
field lines are primarily bended azimuthally by rotation instead of radially by infall.}
\label{Fig:sketch}
\end{figure*}

Along the pseudo-disc plane, where the drift velocities induced by the magnetic field 
bending are the largest in the envelope, the $r$- and $\phi$- components of 
$\bmath{\varv}_{\rm d}$ can be expanded, using Eq.~\ref{Eq:v_ADr}--\ref{Eq:v_Hphi}, as,
\begin{equation}
\label{Eq:v_dr}
\varv_{{\rm d},r} = \varv_{{\rm H},r} + \varv_{{\rm AD},r} \approx {\eta_{\rm H} \over B} {\partial B_\phi \over \partial z} + {\eta_{\rm AD} B_z \over B^2} {\partial B_r \over \partial z}~,
\end{equation}
and
\begin{equation}
\label{Eq:v_dphi}
\varv_{{\rm d},\phi} = \varv_{{\rm H},\phi} + \varv_{{\rm AD},\phi} \approx -{\eta_{\rm H} \over B} {\partial B_r \over \partial z} + {\eta_{\rm AD} B_z \over B^2} {\partial B_\phi \over \partial z}~,
\end{equation}
respectively. The individual terms on the right-hand-side of 
Eq.~\ref{Eq:v_dr}--\ref{Eq:v_dphi} can be either positive or negative, depending 
on the bending direction of the magnetic field. Hence, the ambipolar and Hall drift 
along the pseudo-disc plane can work either {\it cooperatively} or {\it counteractively} 
in drifting the magnetic field in the azimuthal and radial directions. 
To determine the net direction of magnetic field drift in the collapsing 
envelope, we can utilize the fact that the poloidal magnetic field lines are 
preferentially pinched radially inward, and derive the possible scenarios 
as listed below (see illustrations in Fig.~\ref{Fig:sketch}).

For the radial drift $\varv_{{\rm d},r}$, the combined effect of AD and 
Hall effect is to drift the magnetic field radially outward along the 
pseudo-disc plane, in either of the two following scenarios.
\begin{enumerate}
\item (Cooperative) If the magnetic field lines are bended azimuthally (across the 
pseudo-disc mid-plane) towards the {\it same} direction as the azimuthal 
Hall drift $\bmath{\varv}_{{\rm H},\phi}$, the induced radial Hall drift 
$\bmath{\varv}_{{\rm H},r}$ is always along +$r$ (radially outward). 
In this case, both the ambipolar and Hall drift cooperatively diffuse the 
magnetic field radially outward (see left panel of Fig.~\ref{Fig:sketch}). 
\item (Counteractive) If the azimuthal bending of magnetic field lines goes in the 
{\it opposite} direction of the azimuthal Hall drift $\bmath{\varv}_{{\rm H},\phi}$, 
the induced radial Hall drift $\bmath{\varv}_{{\rm H},r}$ points radially inward along 
-$r$ (see right panel of Fig.~\ref{Fig:sketch}). 
Such a configuration normally occurs in the innermost envelope where gas 
rotation becomes large enough to bend the magnetic field azimuthally in its direction. 
However, the outward ambipolar drift generally dominates such an inward Hall drift, 
so that the total radial drift still points outward. 
Basically, the following condition (Eq.~\ref{Eq:BrgtBph}) is satisfied in 
the innermost envelope, as the magnetic field lines are severely pinched radially 
but much less so azimuthally 
(${\partial B_r \over \partial z} \gg {\partial B_\phi \over \partial z}$). 
Assuming a characteristic scale height of variation of magnetic field across the 
pseudo-disc along the $z$-direction, $\varv_{{\rm d},r}>0$ is approximately 
equivalent to,
\begin{equation}
\label{Eq:BrgtBph}
|B_r| > {|\eta_{\rm H}| \over \eta_{\rm AD}} |B_\phi|~,
\end{equation}
in which the ratio ${|\eta_{\rm H}| \over \eta_{\rm AD}}$ is around unity in the 
innermost envelope, depending on the microphysics, especially the grain size 
distribution. 
\end{enumerate}

The azimuthal drift $\varv_{{\rm d},\phi}$ of magnetic fields is usually dominated 
by the Hall component $\varv_{{\rm H},\phi}$, because the azimuthal bending of 
magnetic fields that determines $\varv_{{\rm AD},\phi}$ is only minor in the 
envelope. Since the ambipolar drift $\varv_{{\rm AD},\phi}$ always points away 
from the direction of azimuthal magnetic field bending, we can obtain the 
following two scenarios for the azimuthal drift.
\begin{enumerate}
\item (Cooperative) If the magnetic field lines are bended azimuthally in the 
{\it opposite} direction to the azimuthal Hall drift, both the ambipolar 
and Hall drift, $\varv_{{\rm AD},\phi}$ and $\varv_{{\rm H},\phi}$, 
cooperatively weaken the the magnetic field bending in the azimuthal direction 
(see right panel of Fig.~\ref{Fig:sketch}). 
\item (Counteractive) If the magnetic field lines are bended azimuthally in the 
{\it same} direction as the azimuthal Hall drift, ambipolar and Hall drift 
compete with each other (see left panel of Fig.~\ref{Fig:sketch}).
Usually $\varv_{{\rm H},\phi} \gg \varv_{{\rm AD},\phi}$ 
is satisfied in the envelope, i.e., the azimuthal Hall drift dominates 
the total azimuthal drift $\varv_{{\rm d},\phi}$. 
However, if AD in the envelope is already efficient in relaxing the radial 
pinching of magnetic fields (reducing $|{\partial Br \over \partial z}|$), 
the azimuthal Hall drift $\varv_{{\rm H},\phi}$ can be reduced to be comparable 
to $\varv_{{\rm AD},\phi}$; in cases where the magnetic field is weak, the direction 
of the total azimuthal drift can instead be determined by the azimuthal ambipolar 
drift that slightly weakens the azimuthal bending of magnetic fields. 
\end{enumerate}

It is worth noting that, if the ambipolar and Hall drift are cooperative 
in diffusing the magnetic field in one direction, they are counteractive 
in the orthogonal direction. Basically, as ambipolar drift always tends 
to relax the field bending, the Hall drift component that operates against 
the ambipolar drift in one direction is constrained by a negative feedback 
(see also \PaperI) from the Hall drift operating together with the 
ambipolar drift in the orthogonal direction. In other words, the cooperative 
drift weakens the magnetic field bending in that direction, which in turn 
reduces the Hall drift competing with the ambipolar drift in the orthogonal 
direction.

The above analysis is not limited to the collapsing envelope, but can be 
applied to the protostellar (or protoplanetary) disc itself. 
The main difference is that magnetic field lines threading the disc are 
primarily bended azimuthally by gas rotation, while the radial pinching 
is less prominent across the disc mid-plane. Thus for the disc case, in analogy to 
the reasoning of the envelope case, the combined effect of ambipolar and Hall drift 
in the azimuthal direction is to straighten the azimuthal bending of magnetic fields. 
In the radial direction, Hall drift usually dominates the total radial drift in 
most cases because of the relatively severe azimuthal bending of magnetic fields. 
However, if Ohmic dissipation is already efficient within the disc, the degree 
of magnetic field bending may become much less severe and hence both ambipolar 
and Hall drift can be limited (see also \PaperI). 
Nonetheless, as the magnetic field geometry becomes more complicated 
in the presence of different types of instabilities in protoplanetary discs
\citep[e.g.,][]{Gressel+2015,BaiStone2017,Bethune+2017,Suriano+2018}, the basic 
principles here may still help understand the local behavior of magnetic fields. 
In what follows, we mainly focus on the behavior of magnetic fields in the 
collapsing envelope, which is the key to the formation and early evolution of 
protostellar discs.

\section{Simulation Setup}
\label{Chap.IC}

To investigate the interplay of non-ideal MHD effects in a collapsing dense 
core, as well as their impact on disc formation and evolution, we follow 
the same numerical set-up of \PaperI~ and carry out two-dimensional (2D) 
axisymmetric simulations using ZeusTW code \citep{Krasnopolsky+2010}. 
The three non-ideal MHD effects (AD, Hall effect, and Ohmic dissipation) 
are included, with magnetic diffusivities obtained by linearly interpolating 
the tabulated equilibrium chemical network \citep{Zhao+2018b}.

The initial conditions are the same as \PaperI. We briefly summarize the 
relevant parameters. The initial core is spherically shaped, with total mass 
$M_c=1.0~M_{\sun}$ and radius $R_c=10^{17}$~cm~$\approx 6684$~AU that 
corresponds to a uniform density of $\rho_0=4.77 \times 10^{-19}$~g~cm$^{-3}$. 
The core is rotating initially as a solid-body with angular speed 
$\omega_0 = 1 \times 10^{-13}$~s$^{-1}$ that corresponds to a ratio of 
rotational to gravitational energy $\beta_{\rm rot} \approx 0.025$ 
\citep[the typical value from][]{Goodman+1993}. 
The initial magnetic field is either aligned or anti-aligned with the angular 
angular velocity vector ($\bmath{\Omega}$), with a uniform field strength $B_0$ 
of 42.5~$\mu$G for strong field case, 21.3~$\mu$G for weak field case, 
and 10.6~$\mu$G for very weak field case, which gives a dimensionless mass-to-flux 
$\lambda$ ($\equiv {M_{\rm c} \over \pi R_{\rm c}^2 B_0}2\pi\sqrt{G}$) of 2.4, 4.8, 
and 9.6, respectively. We adopt the a spherical coordinate system 
($r$, $\theta$, $\phi$) with non-uniform grid spacing along $r$- 
and $\theta$- directions. The smallest cell size is set to $\delta r=0.2$~AU 
for cells next to the inner boundary $r_{\rm in}=2$~AU. 
Note that the direction of initial rotation is along +$\phi$. 

As in \PaperI, the magnetic diffusivities are computed using the tabulated 
fractional abundances of charged species from \citet{Zhao+2018b}. 
We explore different cosmic-ray (CR) ionization rates of 
$\zeta_0^{\rm H_2}=10^{-17}$~s$^{-1}$ and $10^{-16}$~s$^{-1}$ at the cloud edge 
with a characteristic attenuation length of $\sim$200~g~cm$^{-2}$ \citep{Padovani+2018}. 
We use 20 size bins to model the MRN \citep[Mathis-Rumpl-Nordsieck;][]{Mathis+1977} 
grain size distribution, fixing the power law index at -$3.5$ and the maximum grain 
size at $a_{\rm max}=0.25~\mu$m, but varying the minimum grain size $a_{\rm min}$ = 
0.005, 0.03, and 0.1~$\mu$m, for MRN, and opt3 (optimal for Hall effect), 
and trMRN (optimal for AD) models, respectively (see notes of 
Table~\ref{Tab:model1}).

To avoid intolerably small time steps, we impose relatively small d$t$ floors for 
AD and Hall effect, with d$t_{\rm floor,AD}=1\times10^5$~s and 
d$t_{\rm floor,H}=3\times10^4$~s, 
which cap the ambipolar and Hall diffusivities.\footnote{The cap of 
$\eta_{\rm AD}$ and $\eta_{\rm H}$ is computed for each cell as 
CFL${|\delta x|_{\rm min}^2 \over 4 {\rm d}t_{\rm floor}}$, where CFL 
is the Courant-Friedrichs-Lewy number that is set to 0.4 for AD and 0.2 for 
Hall, and $|\delta x|_{\rm min}$ is the smallest of the cell's sizes along 
$r$ and $\theta$ directions.} 
Similar to \PaperI, we place a resistivity floor for Ohmic dissipation, which 
equals to the smaller of 10$^{18}$~cm$^{2}$~s$^{-1}$ and $\eta_{\rm H}$, to 
ensure the stability of the Hall solver but to not noticeably weaken the electric 
current density in the inner envelope \citep{Krasnopolsky+2011}. Note that the 
Ohmic diffusivity in the disc is mostly above 10$^{20}$~cm$^{2}$~s$^{-1}$ 
(well-above the resistivity floor), which is large enough to limit the radial 
($B_r$) and azimuthal ($B_\phi$) components of the magnetic field as well as 
the corresponding AD and Hall effect in the disc. 

We summarize a total of 32 numerical models in Table~\ref{Tab:model1}--\ref{Tab:model3}, 
surveying the parameter space of magnetic field strength and direction, grain size 
distribution, and CR ionization rate. 
\begin{table}
\caption{Model Parameters for strong B-field $B_0\approx42.5~\mu$G ($\lambda$$\sim$2.4)}
\label{Tab:model1}
\resizebox{1.1\columnwidth}{!}{
\begin{tabular}{lcccc}
\hline\hline
Model$^\ddagger$ & Grain Size$^\dagger$ & $\zeta_0^{\rm H_2}$ & $\beta_{\rm rot}$ & Radius \& Morphology$^\ast$ \\
& Dist. & (10$^{-17}$~s$^{-1}$) & & (AU) \\
\hline
2.4MRN\_AH$^-$O & MRN & 1 & 0.025 & <2 \\
2.4MRN\_AH$^+$O & MRN & 1 & 0.025 & <2 \\
\hline
2.4min1\_AH$^-$O & min1 & 1 & 0.025 & $\sim$20 (Disc+Spiral/Ring) \\
2.4min1\_AH$^+$O & min1 & 1 & 0.025 & $\downarrow$<2 $\Rightarrow$ $\lesssim$20$^\circlearrowright$ \\
\hline
2.4opt3\_AH$^-$O & opt3 & 1 & 0.025 & 20--30 (Disc+Spiral/Ring) \\
2.4opt3\_AH$^+$O & opt3 & 1 & 0.025 & $\downarrow$<2 $\Rightarrow$ $\sim$30$^\circlearrowright$ \\
2.4$_{\rm Slw}$opt3\_AH$^-$O & opt3 & 1 & 6.25$\times$$10^{-3}$ & $\lesssim$30 \\
2.4$_{\rm Slw}$opt3\_AH$^+$O & opt3 & 1 & 6.25$\times$$10^{-3}$ & $\downarrow$<2 $\Rightarrow$ $\sim$30$^\circlearrowright$ \\
2.4$_{\rm NoRot}$opt3\_AHO & opt3 & 1 & 0 & $\sim$30$^\circlearrowright$ \\
\hline
2.4trMRN\_AO & trMRN & 1 & 0.025 & $\lesssim$20 (Disc+Spiral/Ring) \\
2.4trMRN\_AH$^-$O & trMRN & 1 & 0.025 & $\sim$20 (Disc+Spiral/Ring) \\
2.4trMRN\_AH$^+$O & trMRN & 1 & 0.025 & $\lesssim$20 (Disc+Spiral/Ring) \\
2.4$_{\rm NoRot}$trMRN\_AHO & trMRN & 1 & 0 & --- \\
\hline
2.4CR10opt3\_AH$^-$O & opt3 & 10 & 0.025 & $\lesssim$10 \\
2.4CR10opt3\_AH$^+$O & opt3 & 10 & 0.025 & $\downarrow$<2 $\Rightarrow$ $\sim$10$^\circlearrowright$ \\
2.4CR50opt2\_AH$^-$O & opt2 & 50 & 0.025 & <2 \\
2.4CR50opt2\_AH$^+$O & opt2 & 50 & 0.025 & <2 \\
\hline\hline
\end{tabular}
}
\\
$\dagger$~MRN: full MRN distribution with $a_{\rm min}$=0.005~$\mu$m \\
$\dagger$~min1: truncated MRN distribution with $a_{\rm min}$=0.01~$\mu$m \\
$\dagger$~opt3: truncated MRN distribution with $a_{\rm min}$=0.03~$\mu$m, 
with which Hall diffusivity reaches an optimal level in the inner envelope \\
$\dagger$~trMRN: truncated MRN distribution with $a_{\rm min}$=0.1~$\mu$m \\
$\dagger$~LG: singly-sized grains with $a$=1.0~$\mu$m; note that LG models have $\eta_{\rm H}$>0 at the envelope scale, the opposite to other size distributions \\
$\ddagger$~AH$^-$O: AD+Hall+Ohmic model with anti-aligned configuration ($\bmath{\Omega \cdot B}<0$) \\
$\ddagger$~AH$^+$O: AD+Hall+Ohmic model with aligned configuration ($\bmath{\Omega \cdot B}>0$) \\
$\ddagger$~AO: AD+Ohmic model \\
$\ddagger$~$_{\rm Slw}$: model with slow initial core rotation \\
$\ddagger$~$_{\rm NoRot}$: model with zero initial core rotation \\
$\ast$~The $\uparrow$ or $\downarrow$ symbol indicates that the disc radius is growing or shrinking, repectively \\
$\ast$~The $^\circlearrowright$ symbol indicates that the disc is counter-rotating with respect to the initial core rotation \\
\end{table}

\begin{table}
\caption{Model Parameters for weak B-field $B_0\approx21.3~\mu$G ($\lambda$$\sim$4.8)}
\label{Tab:model2}
\resizebox{1.1\columnwidth}{!}{
\begin{tabular}{lcccc}
\hline\hline
Model & Grain Size & $\zeta_0^{\rm H_2}$ & $\beta_{\rm rot}$ & Radius \& Morphology \\
& Dist. & (10$^{-17}$~s$^{-1}$) & & (AU) \\
\hline
4.8MRN\_AH$^-$O & MRN & 1 & 0.025 & $\sim$13$\downarrow$<2 \\
4.8MRN\_AH$^+$O & MRN & 1 & 0.025 & $\sim$12$\downarrow$<2 \\
\hline
4.8min1\_AH$^-$O & min1 & 1 & 0.025 & $\sim$25 (Disc+Spiral/Ring) \\
4.8min1\_AH$^+$O & min1 & 1 & 0.025 & $\downarrow$<2 $\Rightarrow$ $\lesssim$23$^\circlearrowright$ \\
\hline
4.8opt3\_AH$^-$O & opt3 & 1 & 0.025 & 20--30 (Disc+Spiral/Ring) \\
4.8opt3\_AH$^+$O & opt3 & 1 & 0.025 & $\sim$15$\downarrow$<2 $\Rightarrow$ $\lesssim$20$^\circlearrowright$ \\
4.8$_{\rm Slw}$opt3\_AH$^-$O & opt3 & 1 & 6.25$\times$$10^{-3}$ & $\lesssim$30 \\
4.8$_{\rm Slw}$opt3\_AH$^+$O & opt3 & 1 & 6.25$\times$$10^{-3}$ & $\downarrow$<2 $\Rightarrow$ $\lesssim$20$^\circlearrowright$ \\
4.8$_{\rm NoRot}$opt3\_AHO & opt3 & 1 & 0 & $\sim$35$^\circlearrowright$ \\
\hline
4.8trMRN\_AO & trMRN & 1 & 0.025 & $\lesssim$30 (Disc+Spiral/Ring) \\
4.8trMRN\_AH$^-$O & trMRN & 1 & 0.025 & $\sim$30 (Disc+Spiral/Ring) \\
4.8trMRN\_AH$^+$O & trMRN & 1 & 0.025 & $\lesssim$30 (Disc+Spiral/Ring) \\
4.8$_{\rm NoRot}$trMRN\_AHO & trMRN & 1 & 0 & --- \\
\hline
4.8CR10opt3\_AH$^-$O & opt3 & 10 & 0.025 & $\sim$16$\downarrow$$\sim$10 \\
4.8CR10opt3\_AH$^+$O & opt3 & 10 & 0.025 & $\sim$12$\downarrow$<2 $\Rightarrow$ $\sim$12$^\circlearrowright$ \\
4.8CR50opt2\_AH$^-$O & opt2 & 50 & 0.025 & $\lesssim$10$\downarrow$<2 \\
4.8CR50opt2\_AH$^+$O & opt2 & 50 & 0.025 & $\lesssim$8$\downarrow$<2 \\
\hline\hline
\end{tabular}
}
\end{table}

\begin{table}
\caption{Model Parameters for very weak B-field $B_0\approx10.6~\mu$G ($\lambda$$\sim$9.6)}
\label{Tab:model3}
\resizebox{1.1\columnwidth}{!}{
\begin{tabular}{lcccc}
\hline\hline
Model & Grain Size & $\zeta_0^{\rm H_2}$ & $\beta_{\rm rot}$ & Radius \& Morphology \\
& Dist. & (10$^{-17}$~s$^{-1}$) & & (AU) \\
\hline
9.6opt3\_AH$^-$O & opt3 & 1 & 0.025 & 20--30 (Disc+Spiral/Ring) \\
9.6opt3\_AH$^+$O & opt3 & 1 & 0.025 & $\sim$40 (Disc+Spiral/Ring) \\
9.6opt3\_AHO & opt3 & 1 & 0 & --- \\
\hline\hline
\end{tabular}
}
\end{table}

\section{Simulation Results}
\label{Chap.SimulResult}

As shown in Table~\ref{Tab:model1}--\ref{Tab:model2}, disc formation is sensitive to 
both the grain size distribution and the CR ionization rate. 
A high $\zeta_0^{\rm H_2}$ (at the core scale) close to $5 \times 10^{-16}$~s$^{-1}$, 
or the inclusion of the large population of VSGs (e.g., MRN models) strongly suppresses 
the formation of sizable RSDs larger than the $\sim$2~AU inner boundary. Similar to 
\PaperI, the polarity of the magnetic field does not determine whether discs form or not, 
but can affect the direction of disc rotation in relatively strongly magnetized cores 
($\lambda \lesssim 5$). However, in comparison to \PaperI~ (with no AD), 
counter-rotating discs only appear in the aligned ($\bmath{\Omega \cdot B}>0$) models 
with relatively small $a_{\rm min}$, but not in the aligned models with $a_{\rm min}$ 
increased to $\sim$0.1~$\mu$m (or above); it is a phenomenon controlled by the relative 
magnitude of $\eta_{\rm AD}$ and $\eta_{\rm H}$, and will be discussed in details 
in \S~\ref{S.HallRegime} and ~\ref{S.ADRegime}. 

Note that the disc radii are on the order of 20--30~AU in models with the 
canonical CR ionization rate of 10$^{-17}$~s$^{-1}$; many of these discs or 
their extended spiral structures tend to grow further at later times, which can 
be better demonstrated in 3D studies. In contrast, the disc radii in models with 
high CR ionization rate ($\gtrsim$10$^{-16}$~s$^{-1}$) remain at $\lesssim$10~AU 
throughout their evolution.

\subsection{Suppression of Disc Formation by VSGs}
\label{S.MRN}

\begin{figure}
\includegraphics[width=1.2\columnwidth]{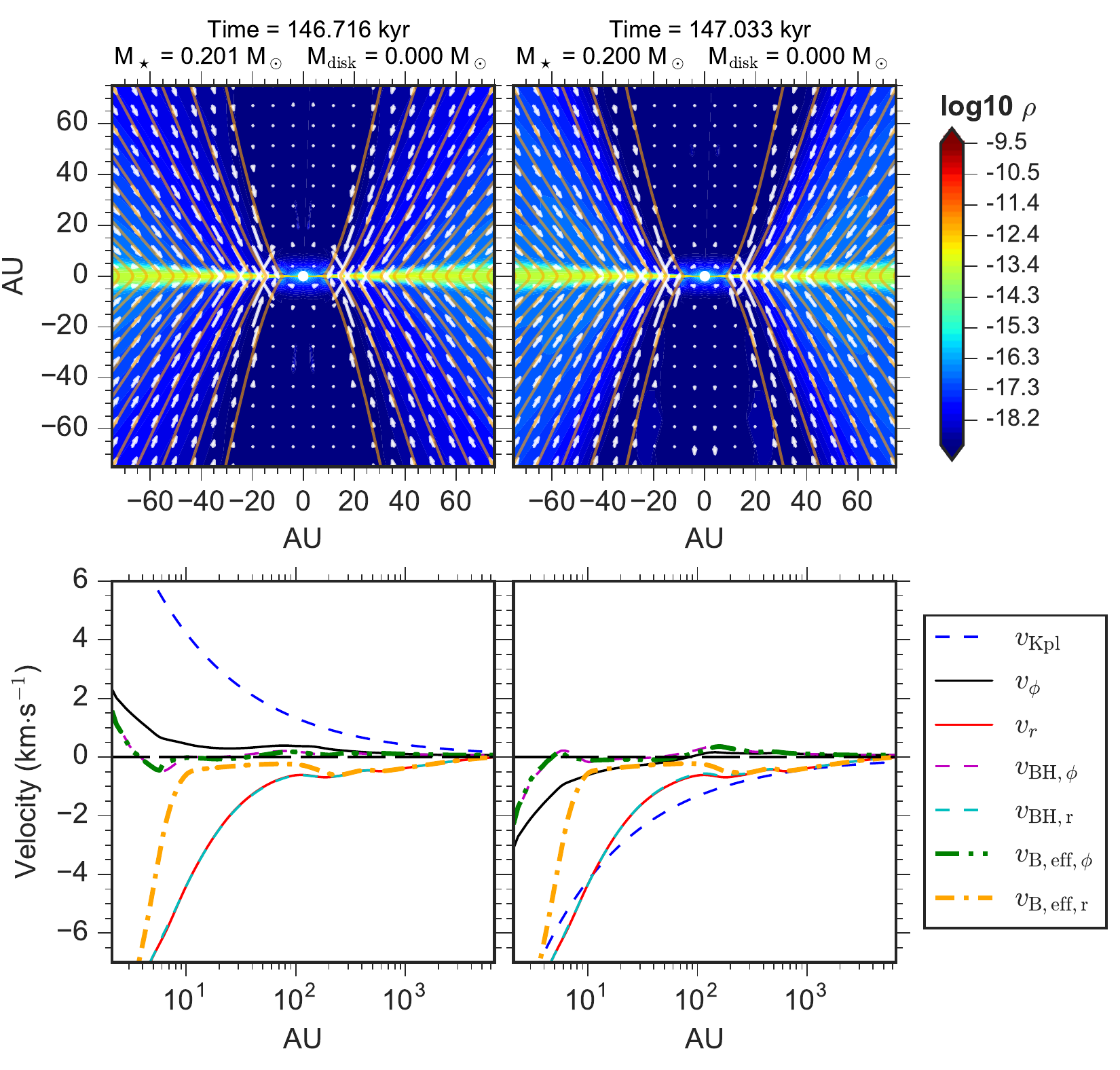}
\begin{tabular}{lcc}
& \hspace*{0.25in} (a) Anti-aligned: $\bmath{\Omega \cdot B}<0$ & \hspace*{0.10in} (b) Aligned: $\bmath{\Omega \cdot B}>0$\\
\end{tabular}
\caption{Mass density distribution (top) and velocity profile along the equator 
(bottom) for model 2.4MRN\_AH$^-$O (left panels) and model 2.4MRN\_AH$^+$O 
(right panels). White arrows and orange lines in the top panel are the velocity field 
vectors and magnetic field lines, respectively.}
\label{Fig:2.4MRN}
\end{figure}

When the standard MRN size distribution is adopted, in which the highly conductive 
VSG population is suppressing the magnetic diffusivities in the collapsing envelope, 
\citep{Zhao+2016,Dzyurkevich+2017,Zhao+2018b}, 
including all three non-ideal MHD effects does not save disc formation, 
which is in agreement with the result of \citet{Li+2011}. 
As shown in Fig.~\ref{Fig:2.4MRN}, the results from the anti-aligned model 
2.4MRN\_AH$^-$O and the aligned model 2.4MRN\_AH$^+$O are nearly identical, 
i.e., both show no obvious RSDs larger than $r_{\rm in}$ (2~AU) throughout the 
protostellar collapse and accretion phase. The main difference is the rotation 
direction of the inner accreting flow ($\lesssim$100~AU) around the central object, 
which is the same as the envelope rotation in the anti-aligned case while the opposite 
to the envelope rotation in the aligned case. However, the rotation speed $\varv_\phi$ 
in both cases are well below the Keplerian speed, consistent with the lack of 
rotationally supported structures in the density distribution.

\begin{figure}
\includegraphics[width=1.0\columnwidth]{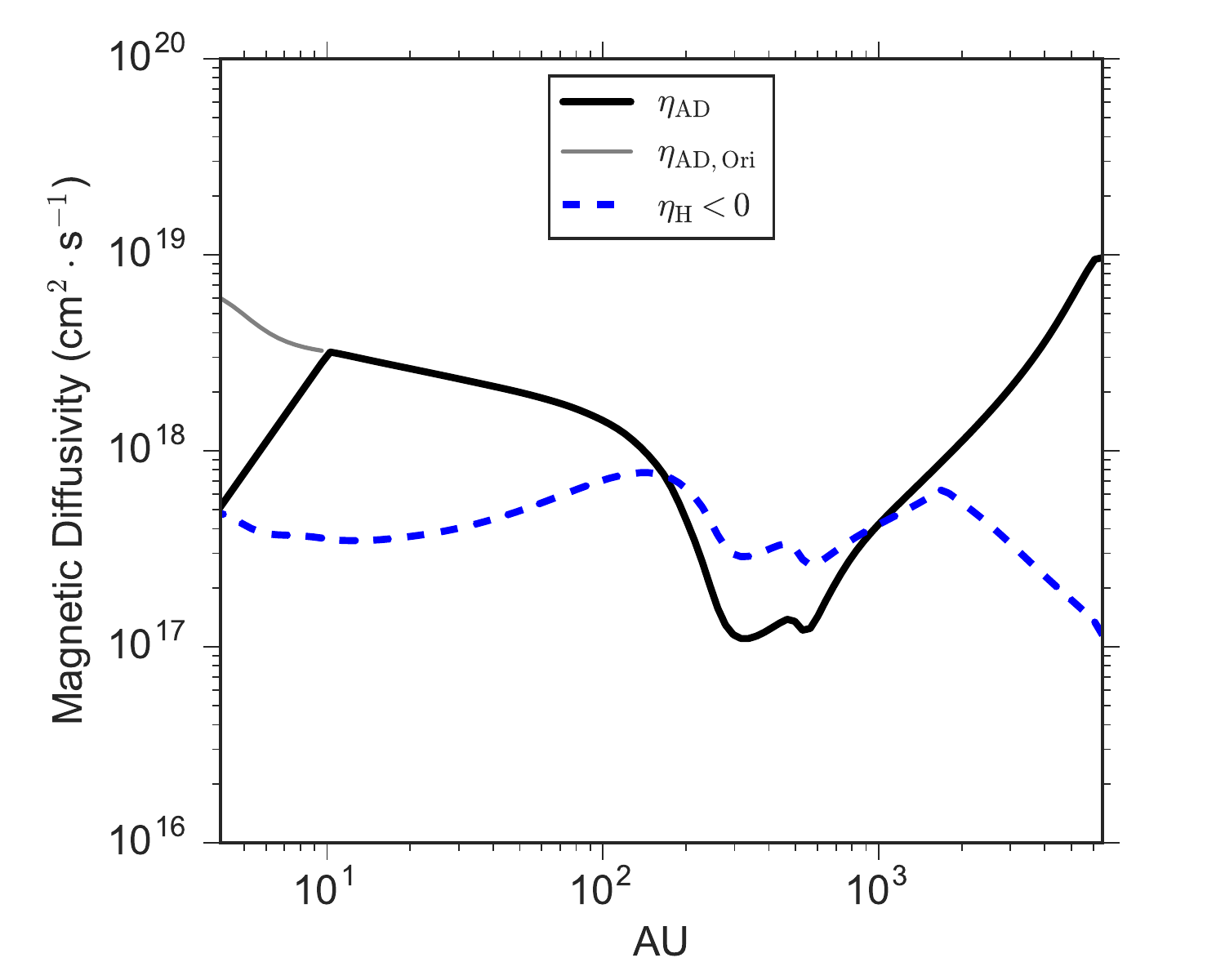}
\caption{Radial profile of the magnetic diffusivities along the equator for the 
2.4MRN\_AH$^-$O model at 146.716~kyr. The grey solid line of $\eta_{\rm AD,Ori}$ 
implies the original ambipolar diffusivity unaffected by the AD d$t$ floor. 
Note that $\eta_{\rm H}$ is low enough in the MRN models that Hall d$t$ floor 
is not triggered.}
\label{Fig:2.4MRNeta}
\end{figure}

The absence of RSDs in the MRN models is a result of the insufficient drift 
of magnetic fields in the bulk envelope. The drift velocity 
$\bmath{\varv}_{\rm d}$ can be estimated in Fig.~\ref{Fig:2.4MRN} 
by the difference between the effective velocity of the magnetic field lines 
$\bmath{\varv}_{\rm B,eff}$ and the bulk neutral velocity $\bmath{\varv}$. 
For convenience of analysis, we also define an effective velocity of the magnetic 
field lines due to Hall drift alone (see also \PaperI), as 
\begin{equation}
\bmath{\varv}_{\rm BH} = \bmath{\varv} + \bmath{\varv}_{\rm H}~.
\label{Eq:v_iH}
\end{equation}
In either model, the azimuthal drift of magnetic fields is primarily dominated by 
the azimuthal Hall drift: $\varv_{{\rm B,eff},\phi} \approx \varv_{{\rm BH},\phi}$; 
while the radial drift of magnetic fields is almost entirely determined by the 
radial ambipolar drift: 
|$\varv_{{\rm B,eff},r}$| $\ll$ |$\varv_r$| $\approx$ |$\varv_{{\rm BH},r}$|. 
However, the radial ambipolar drift only becomes prominent in the inner 
$\lesssim$100~AU, but is nearly vanishing in the outer part of the collapsing 
envelope. In other words, the amount of magnetic flux being dragged into the 
inner $\lesssim$100~AU region is not much different from the ideal MHD limit. 
As already shown in previous studies \citep{Li+2011,Zhao+2018a,Lam+2019}, radial 
diffusion of magnetic fields only within the inner $\lesssim$100~AU by 
AD is unable to save disc formation;\footnote{Note that the inner $\lesssim$100~AU, 
where $\varv_{{\rm B,eff},r}$ drops below 0.5~km~s$^{-1}$ due to ambipolar drift 
($\varv_{{\rm AD},r} \sim$ a few km~s$^{-1}$), corresponds to the so-called 
``diffusion DEMS'' (Decoupling-Enabled Magnetic Structures) demonstrated 
in \citet{Lam+2019} or the AD-induced magnetic plateau found by \citet{Masson+2016}. 
The existence of such structures is a natural diffusive response of the 
magnetic field, instead of a prerequisite for the formation of sizable RSDs.} 
either additional radial diffusion in the bulk envelope \citep[limiting the magnetic 
flux arrived in the inner envelope;][]{Zhao+2018a} 
and/or efficient azimuthal diffusion by Hall effect (limiting the radial current 
$J_r \propto {\partial B_\phi \over \partial z}$; see also \PaperI) are needed 
for weakening the magnetic braking torque, which are unfortunately 
not achieved with the standard MRN size distribution. More specifically 
(Fig.~\ref{Fig:2.4MRNeta}), the ambipolar diffusivity $\eta_{\rm AD}$ along the 
pseudo-disc (equatorial plane) is on the order of $\sim$10$^{17}$~cm$^2$~s$^{-1}$ 
at 10$^2$--10$^3$~AU scale; while the Hall diffusivity $\eta_{\rm H}$ along the 
pseudo-disc remains at a few 10$^{17}$~cm$^2$~s$^{-1}$ in the inner $\lesssim$100~AU. 
Such values of diffusivities in the inner envelope are 1--2 orders of magnitude 
below the required level for disc formation suggested by \citet{Shu+2006} and 
\citet{Krasnopolsky+2010}. Note that in the inner envelope, $\eta_{\rm AD}$ 
only increases above 10$^{18}$~cm$^2$~s$^{-1}$ (a few times larger than $\eta_{\rm H}$) 
within the $\lesssim$100~AU region where ambipolar drift becomes prominent. 

\subsection{Hall Dominated Collapse: Bimodal Disc Rotation}
\label{S.HallRegime}

As we slightly increase $a_{\rm min}$, i.e., truncating off the large population 
of VSGs ($\lesssim$10~nm) from the grain size distribution, disc formation 
becomes insensitive to further changes in grain sizes. 
We first focus on the models adopting $a_{\rm min} = 0.03~\mu$m, with which the 
Hall diffusivity $\eta_{\rm H}$ reaches an optimal level throughout the collapsing 
envelope \citep{Zhao+2018b,Koga+2019}, while the ambipolar diffusivity 
$\eta_{\rm AD}$ has not yet reached its optimal level \citep{Zhao+2016}. 
We show that in such models AD and Hall effect operate together to promote disc formation. 
In particular, Hall effect efficiently regulates the magnetic field drift and gas dynamics 
in the inner envelope, causing the discs to rotate in opposite directions depending on 
the polarity of the magnetic field at the core scale.

\subsubsection{Anti-aligned Case \texorpdfstring{$\bmath{\Omega \cdot B}<0$}{}}
\label{S.HallAnti}

\begin{figure*}
\centerline{\includegraphics[width=1.0\textwidth]{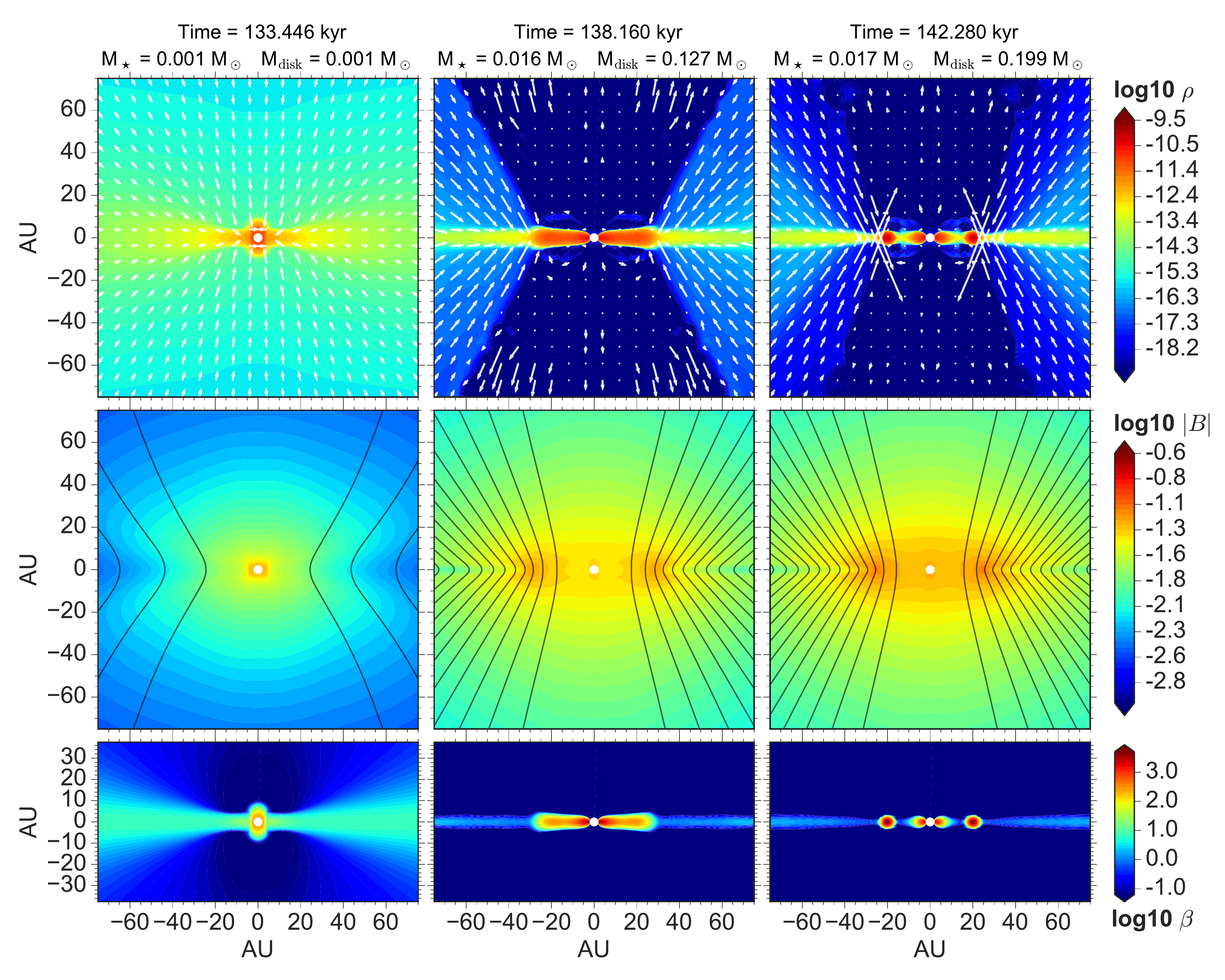}}
\centerline{\includegraphics[width=1.0\textwidth]{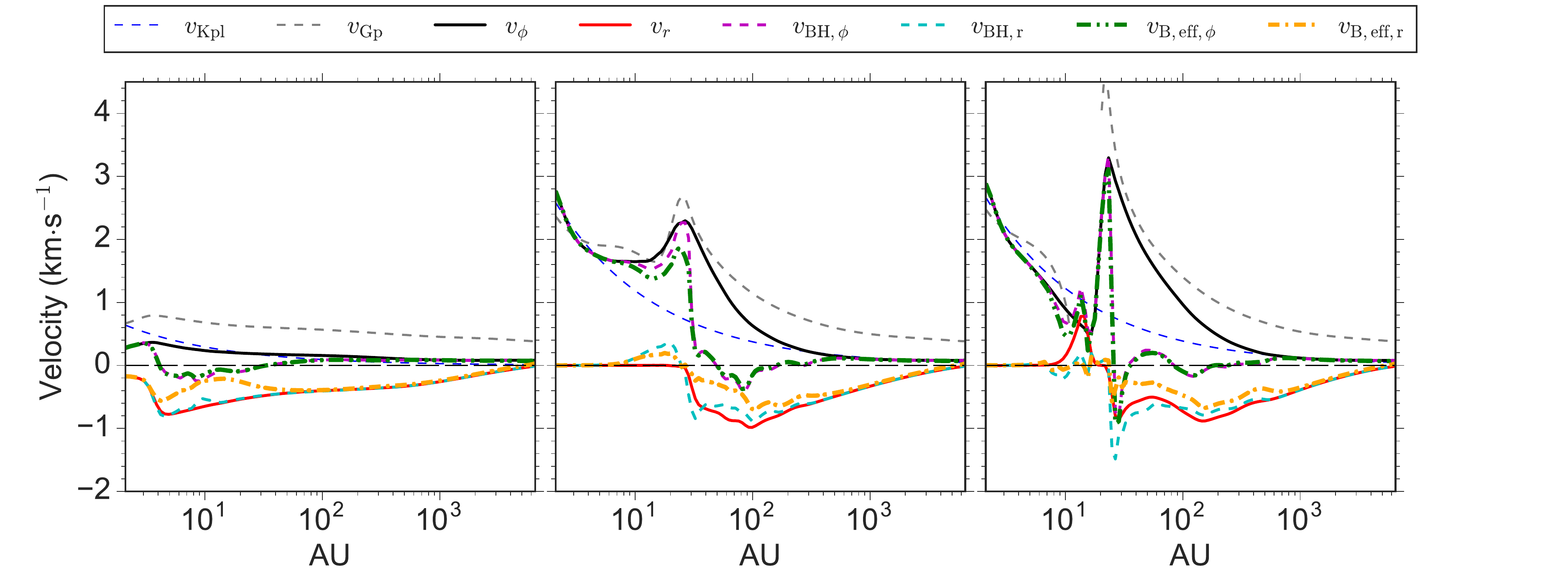}}
\caption{Evolution of disc in the anti-aligned model 2.4opt3\_H$^-$O. First row: 
logarithmic distribution of mass density along with velocity field vectors (white arrows). 
Second row: logarithmic distribution of total magnetic field strength $|B|$ along 
with magnetic field lines (black solid lines). Third row: logarithmic distribution 
of plasma-$\beta$. Fourth row: velocity profile along the equator.}
\label{Fig:2.4opt3AH-}
\end{figure*}

\begin{figure*}
\includegraphics[width=1.0\textwidth]{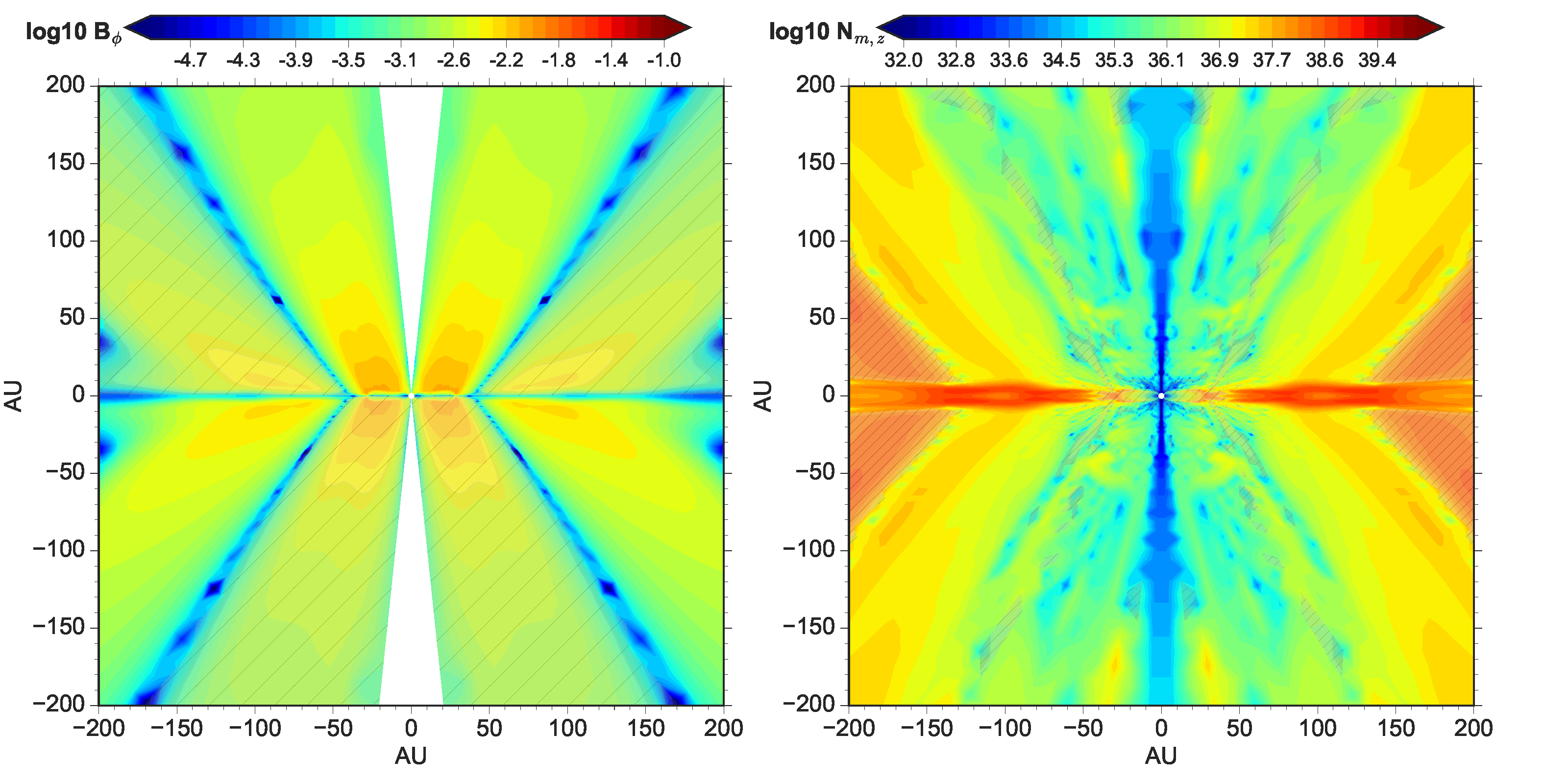}
\caption{Logarithmic distribution of azimuthal magnetic field $B_\phi$ (left panel) and 
magnetic torque $N_{{\rm m},z}$ (right panel) at $t=138.160$~kyr of the anti-aligned 
model 2.4opt3\_AH$^-$O. $B_\phi$ is positive (along +$\phi$) in the unshaded region and 
negative (along -$\phi$) in the shaded region. Similarly, regions of negative magnetic 
torque (along -$z$) are shown as shaded.}
\label{Fig:BphTq58}
\end{figure*}

Fig.~\ref{Fig:2.4opt3AH-} shows the evolution of the inner $\sim$70~AU of the 
anti-aligned model 2.4opt3\_AH$^-$O, in which a RSD forms after the first core 
\citep{Larson1969} stage and grows in radius within $\sim$5~kyrs to $\sim$30~AU. 
The disc formation and growth is a result of the combined effort of AD and Hall 
effect in weakening the magnetic braking torque or exerting a spin-up torque. 
In the azimuthal direction, the magnetic field is predominantly drifted 
towards -$\phi$ along the pseudo-disc by Hall effect 
($\varv_{{\rm BH},\phi} \approx \varv_{{\rm B,eff},\phi}$), like in the 
MRN models but with a more substantial Hall drift velocity $\varv_{{\rm H},\phi}$. 
Similar to \PaperI, $\varv_{{\rm H},\phi}$ here is large enough to over-bend the 
magnetic field lines towards -$\phi$-direction (e.g., Fig.~\ref{Fig:BphTq58}), causing 
a spin-up torque to accelerate the gas rotation along +$\phi$-direction in the inner 
envelope. In the radial direction, both AD and Hall effect contribute to the magnetic 
field drift, with the total drift $\varv_{{\rm d},r}$ almost always pointing towards +$r$ 
(radially outward) in the inner envelope (from the disc edge up to $\lesssim$10$^3$~AU). 
The persistent outward diffusion of magnetic fields decreases the amount of magnetic flux 
that would otherwise be dragged into the innermost region (reducing $B_z$). 
Hence, even as the azimuthal bending of magnetic fields is flipped back to 
+$\phi$-direction in the innermost region by the large gas rotation (along +$\phi$), 
the resulting spin-down (magnetic braking) torque is weakened due to the reduced 
poloidal magnetic field ($B_z$). 

Although the net effect of AD and Hall effect in the radial direction is to move the 
magnetic field radially outward relative to the infalling matter, the two effects can 
either be cooperative or counteractive, depending on the direction of azimuthal 
bending of magnetic fields. For example, at $t=138.160$~kyr (middle panels of 
Fig.~\ref{Fig:2.4opt3AH-}), the radial Hall drift velocity $\varv_{{\rm H},r}$ 
is positive ($\varv_{{\rm BH},r}-\varv_r > 0$) between $\sim$40--400~AU, 
i.e., pointing towards +$r$, which is the same as the direction of the radial 
ambipolar drift $\varv_{{\rm AD},r}$ (as $\varv_{{\rm B,eff},r} - \varv_{{\rm BH},r} > 0$); 
in this case, the radial ambipolar and Hall drift velocities are cooperatively diffusing 
the magnetic field outward. However, between $\sim$27--40~AU, the radial Hall drift 
$\varv_{{\rm H},r}$ flips its sign to negative, i.e., pointing towards -$r$, which 
is the opposite of the radial ambipolar drift $\varv_{{\rm AD},r}$. In spite of such 
a counteractive effect, $\varv_{{\rm AD},r}$ is much larger than $\varv_{{\rm H},r}$ 
(see Eq.~\ref{Eq:BrgtBph} in \S~\ref{S.Interplay}), so that the total radial drift 
still points towards +$r$ (as $\varv_{{\rm d},r}>0$). 
Such an interplay between the radial AD and Hall drift is even more prominent at later 
times (e.g., $t=142.280$~kyr). 

\begin{figure}
\includegraphics[width=1.0\columnwidth]{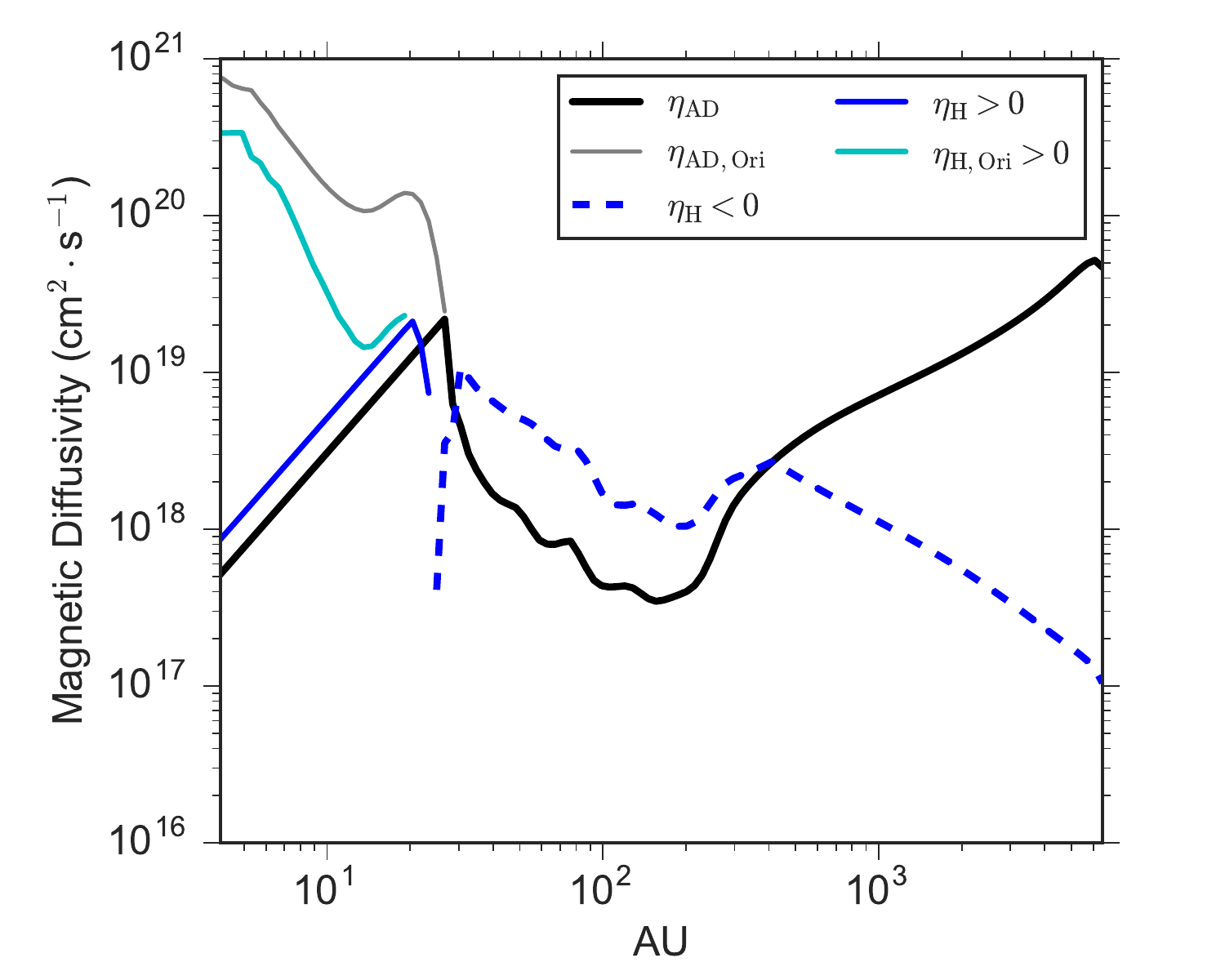}
\caption{Radial profile of the magnetic diffusivities along the equator for the 
2.4opt3\_AH$^-$O model at 138.160~kyr. The grey solid line of $\eta_{\rm AD,Ori}$
implies the original ambipolar diffusivity unaffected by the AD d$t$ floor. 
The cyan solid line of $\eta_{\rm H,Ori}$ implies the original Hall diffusivity 
unaffected by the Hall d$t$ floor.}
\label{Fig:2.4opt3eta}
\end{figure}

In comparison to the MRN model (2.4MRN\_AH$^-$O) above, the elevated role of AD 
and Hall effect in the collapsing envelope of the opt3 model (2.4opt3\_AH$^-$O) 
is primarily caused by the enhanced level of the ambipolar ($\eta_{\rm AD}$) 
and Hall ($\eta_{\rm H}$) diffusivities. 
Comparing Fig.~\ref{Fig:2.4opt3eta} with Fig.~\ref{Fig:2.4MRNeta}, the ambipolar 
diffusivity $\eta_{\rm AD}$ beyond 200~AU scale differs by $\gtrsim$1 order of 
magnitude, with the opt3 model having the larger $\eta_{\rm AD}$ 
($\sim$10$^{18}$--10$^{20}$~cm$^2$~s$^{-1}$ versus 
$\sim$10$^{17}$--10$^{19}$~cm$^2$~s$^{-1}$ in the MRN model). 
As a result, the ambipolar drift in the opt3 model already becomes visible at 
the envelope scale between 200--800~AU ($\varv_{{\rm AD},r}$$\gtrsim$0.1~km~s$^{-1}$), 
as compared to the vanishing $\varv_{{\rm AD},r}$ in the MRN model. 
The Hall diffusivity $\eta_{\rm H}$ in the opt3 model is also larger by 
$\sim$1 order of magnitude than in the MRN model, especially within the 
inner $\lesssim$100--200~AU where Hall effect becomes pronounced in 
regulating the magnetic field and gas dynamics (\ct{Tsukamoto+2017}; \PaperI). 
In fact, between $\sim$40--200~AU of the opt3 model, $\eta_{\rm H}$ is dominating 
over $\eta_{\rm AD}$ by a factor of a few along the equatorial region. 

Under the efficient AD and Hall effect, the RSD formed in the 2.4opt3\_AH$^-$O model 
survives throughout the simulated protostellar phase ($\sim$20~kyr). 
The RSD accretes mass more rapidly from the collapsing envelope than it is 
able to transfer to the central stellar object,\footnote{Initial RSDs formed in 
collapse simulations are generally massive, with moderate dependence on 
the sink treatment \citep{Hennebelle+2020}. In this study, the disc to 
stellar mass ratio is further amplified by the 2D axisymmetric set-up that 
prevents mass accretion to the central stellar object via gravitational torques.} 
and becomes self-gravitating. Ring-like structure \citep[similar to][]{Zhao+2016} 
with large plasma-$\beta$ (the ratio of thermal to magnetic pressure) of 
$\sim$10$^3$--10$^4$ develops after $\sim$8~kyr of the first core formation, 
which would instead be an extended spiral structures in 3D simulations \citep{Zhao+2018a}. 
Because of the low stellar mass, the gas rotation speed is mostly super-Keplerian, 
while remains gravitationally bound with 
$\varv_\phi \lesssim \varv_{\rm Gp} \equiv \sqrt{r {\partial \Phi \over \partial r}}$ 
($\Phi$ is the gravitational potential at radius $r$). 
Note that in this anti-aligned ($\bmath{\Omega \cdot B}<0$) model, the disc rotation 
remains in the same direction as the initial core rotation, because the Hall effect in 
the inner envelope is to spin up the rotation of the accreting flow.

\subsubsection{Aligned Case \texorpdfstring{$\bmath{\Omega \cdot B}>0$}{}}
\label{S.HallAlign}

\begin{figure*}
\centerline{\includegraphics[width=1.0\textwidth]{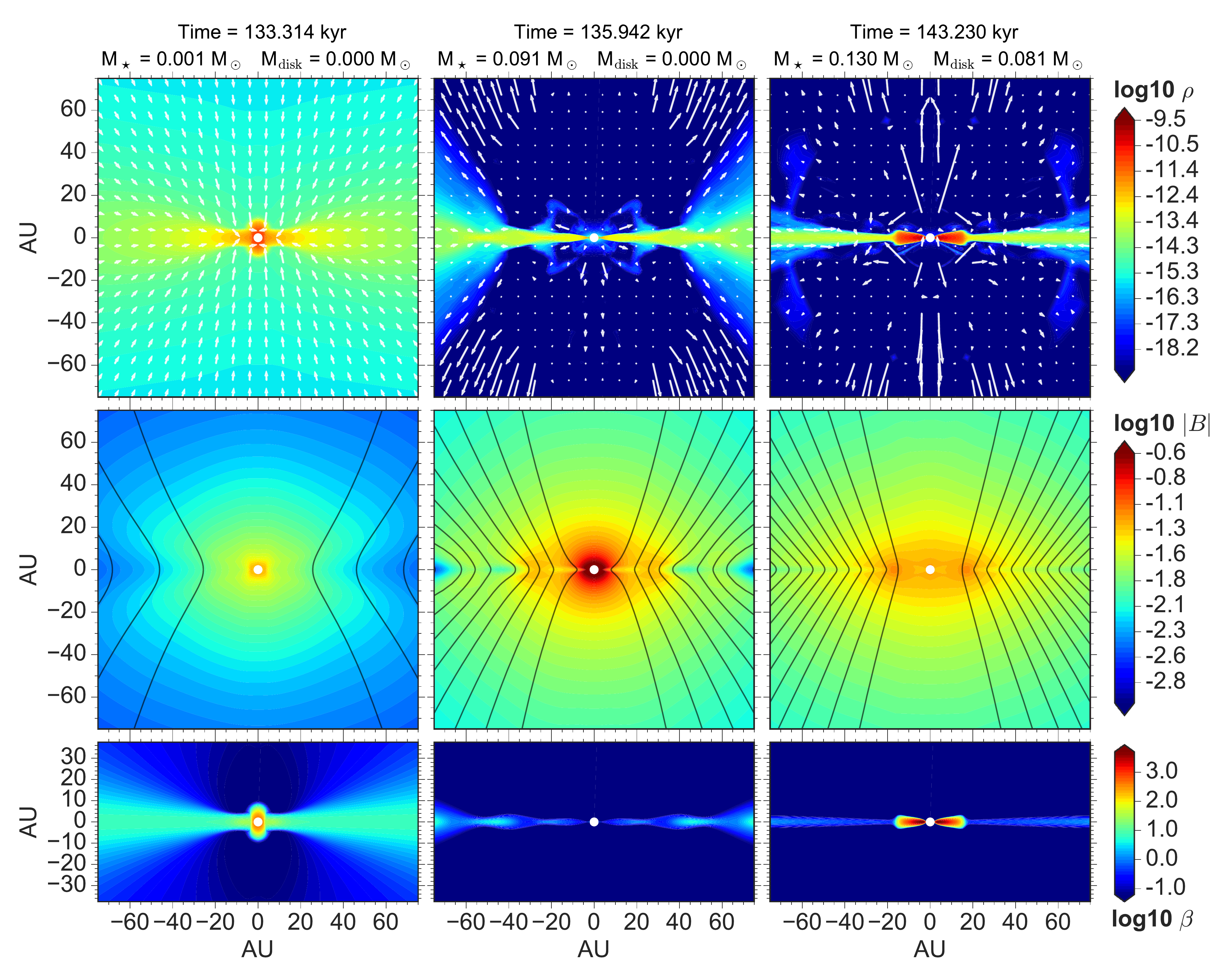}}
\centerline{\includegraphics[width=1.0\textwidth]{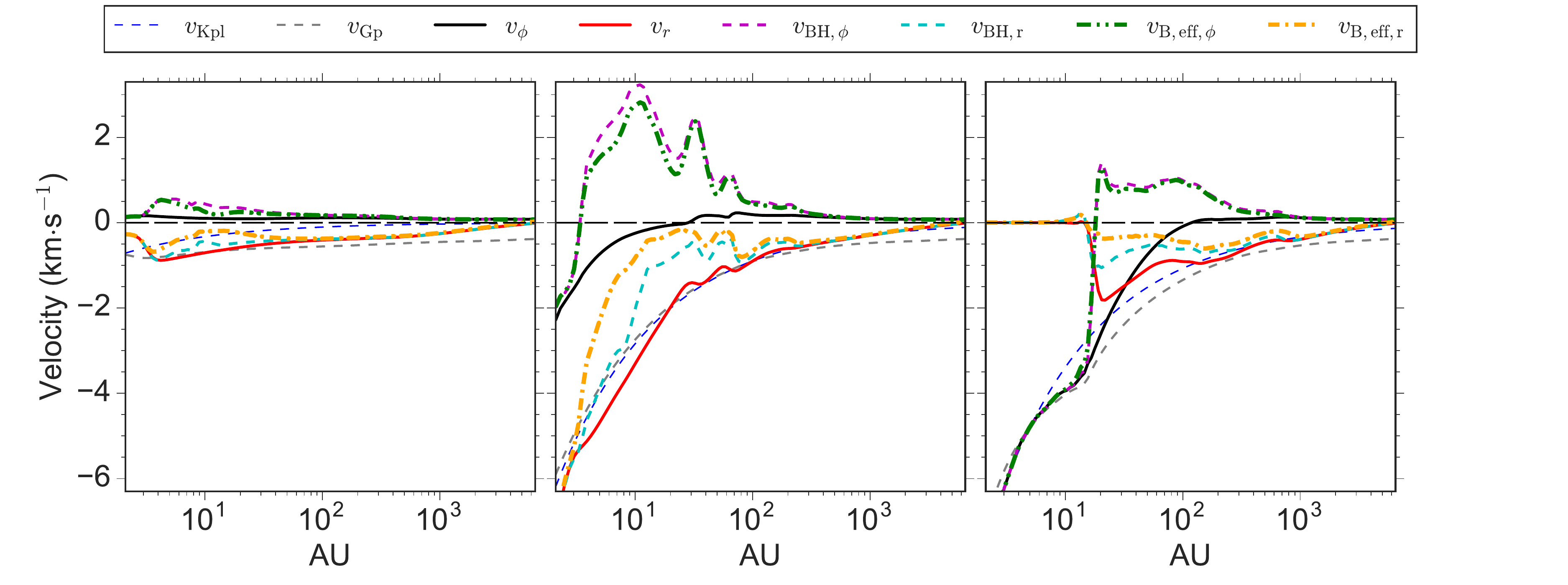}}
\caption{Evolution of disc in the aligned model 2.4opt3-H$^+$O. First row: 
logarithmic distribution of mass density along with velocity field vectors (white arrows). 
Second row: logarithmic distribution of total magnetic field strength $|B|$ along 
with magnetic field lines (black solid lines). Third row: logarithmic distribution 
of plasma-$\beta$. Fourth row: velocity profile along the equator.}
\label{Fig:2.4opt3AH+}
\end{figure*}

We now turn to the aligned model 2.4opt3\_AH$^+$O; the only difference in set-up 
from the 2.4opt3\_AH$^-$O model above is the polarity of the magnetic field. 
As shown in Fig.~\ref{Fig:2.4opt3AH+}, disc formation is initially suppressed 
during the first $\sim$4~kyr after the first core formation, followed by the formation 
of a counter-rotating disc of $\sim$20~AU radius. The process is largely 
similar to the early evolution of the aligned model demonstrated in \PaperI~ 
(see their Fig. 12) where only Hall and Ohmic are included. 
Basically, the azimuthal Hall drift enhances the bending of magnetic fields 
towards the direction of the initial rotation +$\phi$, which not only brakes 
the gas rotation along +$\phi$, but also reverses its direction to -$\phi$ within 
the inner $\sim$200~AU. The azimuthal Hall drift velocity $\varv_{{\rm H},\phi}$ 
(indicated by $\varv_{{\rm BH},\phi}-\varv_\phi$) reaches values as large as 
$\sim$2--3~km~s$^{-1}$ in the inner tens of AU, dominating the total azimuthal drift 
of magnetic fields $\varv_{{\rm d},\phi}$ ($\equiv \varv_{{\rm B,eff},\phi}-\varv_\phi$). 
Although the azimuthal ambipolar drift $\varv_{{\rm AD},\phi}$ is vanishing, it is 
still directed towards -$\phi$ (as $\varv_{{\rm B,eff},\phi}-\varv_{{\rm BH},\phi}<0$), 
which is the opposite to the direction of the azimuthal Hall drift and hence tends to 
slightly relax the azimuthal bending of magnetic fields. 
In the radial direction, both ambipolar and Hall drift are directed along +$r$, 
working together to move the magnetic field radially outward relative to the 
infalling matter within $\lesssim$10$^3$~AU. Because of the enhanced azimuthal bending 
of magnetic fields by the azimuthal Hall drift in the aligned case, the induced radial 
Hall drift $\varv_{\rm H,r}$ is also somewhat larger than the radial ambipolar drift 
$\varv_{{\rm AD},r}$ (in contrast to the anti-aligned case 
where $\varv_{{\rm AD},r}>\varv_{{\rm H},r}$; \S~\ref{S.HallAnti}). 

It is worth noting that the counter-rotating RSD formed in the aligned case does 
not quickly become self-gravitating, and the disc mass remains below the stellar 
mass ($\sim$0.13~$M_{\sun}$) for as long as $\sim$10~kyr (after the disc formation). 
The low disc to stellar mass ratio is primarily because the central stellar object 
has already gathered sufficient mass during the initial disc suppression phase when 
the infalling gas is accreted easily into the inner boundary. However, as the 
counter-rotating RSD develops, most of the infalling gas lands on the disc instead, 
and the stellar accretion slows down greatly (especially in the current 2D set-up). 
Because of the substantial stellar mass, the counter-rotating RSD steadily 
grows in radius to $\sim$30~AU without condensing into ring-like structures like 
in the anti-aligned case (Fig.~\ref{Fig:2.4opt3AH+}). 
Again, the plasma-$\beta$ along the disc mid-plane reaches $\sim$10$^3$--10$^4$.

\subsubsection{Envelope Rotation}
\label{S.Envelope}

\begin{figure*}
\includegraphics[width=1.0\textwidth]{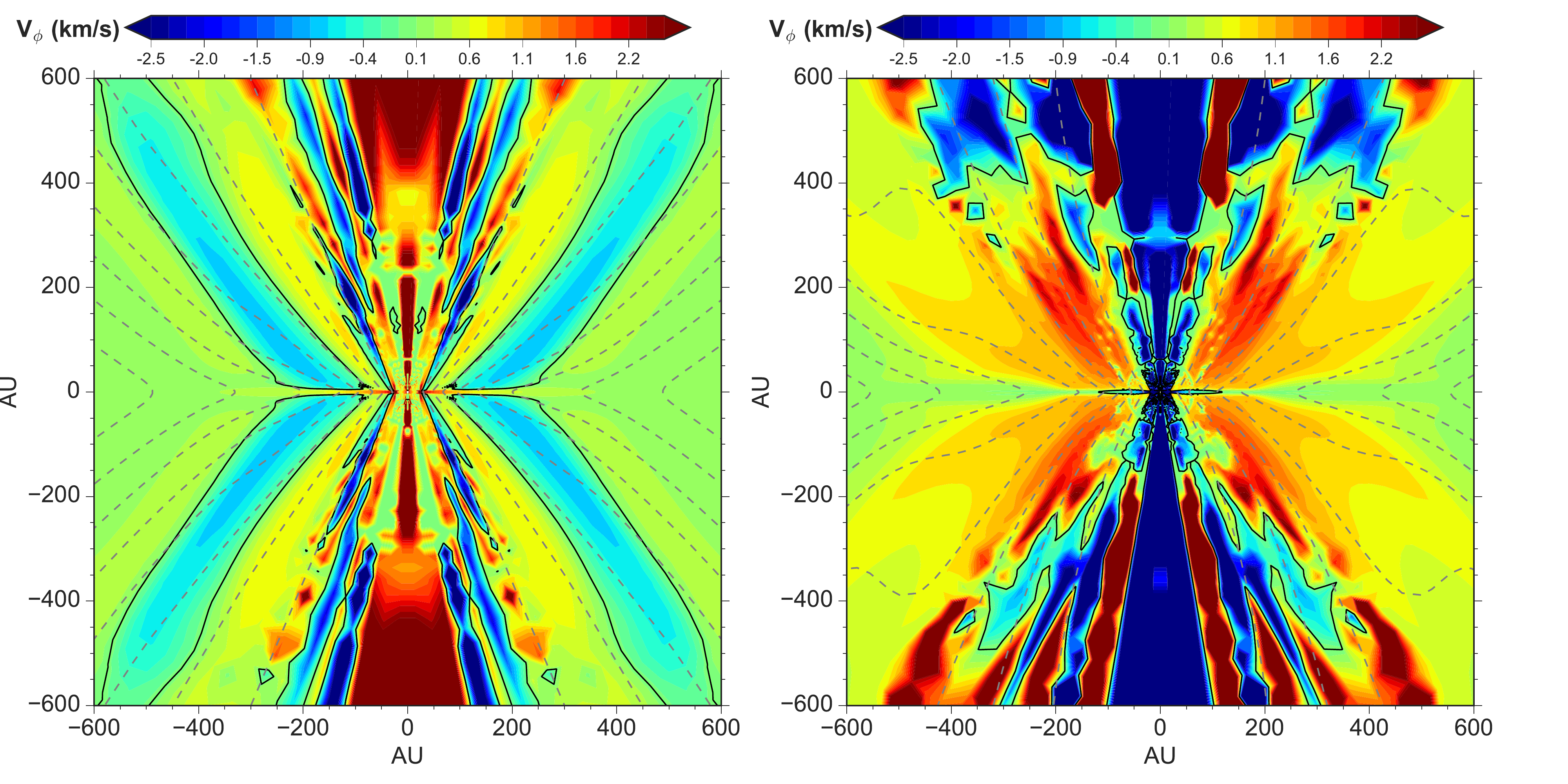}
\caption{Distribution of azimuthal velocity $\varv_\phi$ at 600~AU scale 
envelope for both the anti-aligned model 2.4opt3\_AH$^-$O at $t=142.280$~kyr 
(left panel) and the aligned model 2.4opt3\_AH$^+$O at $t=143.230$~kyr (right panel). 
Negative $\varv_\phi$ values represent rotation motion along -$\phi$ direction. 
Solid contour lines mark positions with $\varv_\phi=0$, where the transition between 
positive and negative $\varv_\phi$ occurs. Grey dashed curves trace the magnetic field 
lines.}
\label{Fig:envelopVph}
\end{figure*}

As demonstrated in \PaperI~ \citep[see also][]{Krasnopolsky+2011,Li+2011,Tsukamoto+2015b}, 
Hall effect can induce counter-rotating motions at the envelope scale by regulating the 
topology of the magnetic field; either a ``butterfly-shaped'' thin shell is counter-rotating 
(anti-aligned case) or the inner envelope enclosing the disc and outflow region is 
counter-rotating (aligned case). Similar structures are present in both opt3 models 
here (2.4opt3\_AH$^-$O and 2.4opt3\_AH$^+$O), in which, despite the inclusion of 
AD as compared to \PaperI~ without AD, Hall effect still dominates over AD in 
the inner envelope due to the choice of $a_{\rm min}=0.03~\mu$m (rendering 
$\eta_{\rm H}>\eta_{\rm AD}$; Fig.~\ref{Fig:2.4opt3eta}). 
As shown in Fig.~\ref{Fig:envelopVph}, in the anti-aligned case, $\varv_\phi$ becomes 
negative (along -$\phi$) within a thin shell between $\sim$100--1000~AU scale; 
while in the aligned case, $\varv_\phi$ is only negative along the inner $\lesssim$100~AU 
equatorial region and along narrow stripes in the bipolar cavity. The extent of 
the counter-rotating region in the anti-aligned case is similar to the Hall 
models in \PaperI~ (see their Figure~10); for the aligned case, the counter-rotating 
regions are somewhat smaller and less prominent comparing to the Hall models in \PaperI~ 
in which the inner $\lesssim$200~AU equatorial region and almost the entire bipolar 
cavity are counter-rotating (see their Figure~14). 

The presence of counter-rotating regions in these Hall dominated cases is a sign of 
angular momentum redistribution among different parts of the collapsing envelope 
by Hall effect. In the anti-aligned case, the negative angular momentum of the 
counter-rotating shells spares the angular momentum budget needed for spinning 
up the gas rotation in the inner envelope and disc. In the aligned case, there is 
an excess of angular momentum at a few 10$^2$~AU scale (reddish ``butterfly-shaped'' 
lobes in the right panel of Fig~\ref{Fig:envelopVph}), which compensates the negative 
angular momentum in the counter-rotating inner equatorial and outflow region 
(see detailed discussions in \PaperI). 
In either case, the deceleration or acceleration of gas rotation in the 
``butterfly-shaped'' envelope region is directed by the magnetic tension force in 
$\phi$-direction, which is regulated by the azimuthal Hall drift $\varv_{{\rm H},\phi}$ 
that points to +$\phi$ in the anti-aligned case and -$\phi$ in the aligned case. 
These preferred directions of azimuthal Hall drift in the ``butterfly-shaped'' regions 
are naturally derived from the convex-shaped (bending towards the equator) poloidal 
magnetic fields therein (Fig.~\ref{Fig:envelopVph}; see also \PaperI). 
Therefore, by choosing $a_{\rm min}=0.03~\mu$m, Hall effect remains sufficient 
at the envelope scale to leave imprints on the angular momentum redistribution.
However, it is no longer the case if a larger $a_{\rm min}$ is adopted, as we will 
discuss next in \S~\ref{S.ADRegime}.

\subsection{AD Dominated Collapse: Unimodal Disc Rotation}
\label{S.ADRegime}

As magnetic diffusivities are sensitive to the grain size distribution 
\citep{Zhao+2016,Dzyurkevich+2017}, we explore the trMRN models adopting 
$a_{\rm min}=0.1~\mu$m, with which the ambipolar diffusivity $\eta_{\rm AD}$ 
reaches an optimal level throughout the collapsing envelope \citep{Zhao+2016}, 
while the Hall diffusivity $\eta_{\rm H}$ drops by $\sim$1 order of magnitude 
below the optimal level of the opt3 models at densities $\lesssim$10$^9$~cm$^{-3}$ 
\citep[see detailed discussions in][]{Zhao+2018b,Koga+2019}. Hence, in these 
trMRN models, AD dominates the diffusion of magnetic fields, while Hall effect 
plays a minor role and has little impact on the direction of gas rotation. 
As we show below, disc rotates along the same direction as the intial core, 
regardless of the polarity of the magnetic field.

\begin{figure*}
\centerline{\includegraphics[width=1.0\textwidth]{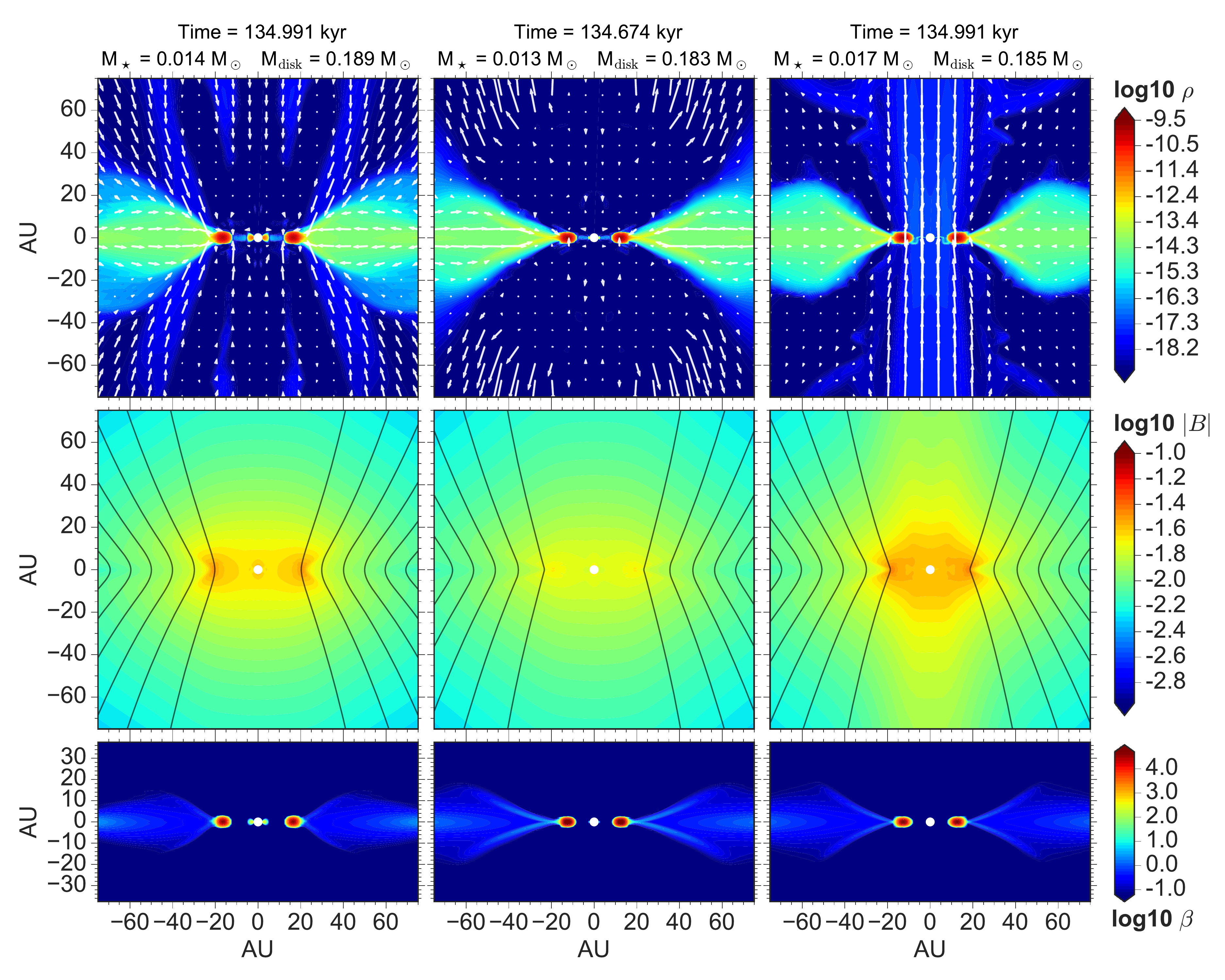}}
\centerline{\includegraphics[width=1.0\textwidth]{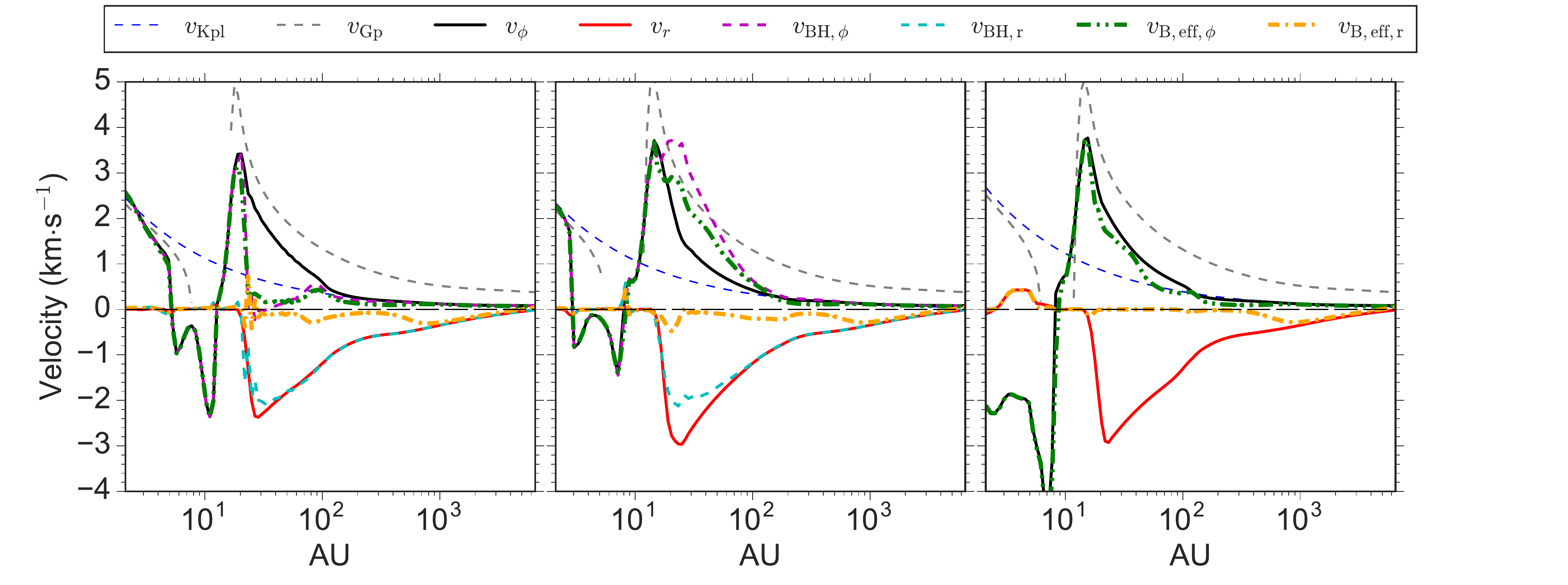}}
\begin{tabular}{lccc}
& \hspace*{-0.8in} (a) \textbf{Anti-aligned}: $\bmath{\Omega \cdot B}<0$ & \hspace*{0.6in} (b) \textbf{Aligned}: $\bmath{\Omega \cdot B}>0$ & \hspace*{0.8in} (c) \textbf{AD model} \\
\end{tabular}
\caption{Disc formation in the trMRN models: anti-align case (left panels), aligned case 
(middle panels), and case with only AD and Ohmic (right panels). 
First row: logarithmic distribution of mass density along with velocity field vectors 
(white arrows). 
Second row: logarithmic distribution of total magnetic field strength $|B|$ along with 
magnetic field lines (black solid lines). 
Third row: logarithmic distribution of plasma-$\beta$. 
Fourth row: velocity profile along the equator.}
\label{Fig:2.4dust1kAH}
\end{figure*}

Fig.~\ref{Fig:2.4dust1kAH} presents three trMRN models: the anti-aligned model 
2.4trMRN\_AH$^-$O, aligned model 2.4trMRN\_AH$^+$O, and the comparison model 2.4trMRN\_AO 
(without Hall effect), at a similar evolution stage when the total mass of the star 
and disc reaches $\sim$0.2~$M_{\sun}$. The disc morphology in the three models is 
nearly identical, showing as self-gravitating ring structures of $\sim$20~AU radius 
\citep[that would be spirals or multiples in full 3D simulations;][]{Zhao+2018a}. 
The gas rotation speed $\varv_\phi$ is positive (along +$\phi$) throughout the 
infalling envelope in all three models, and accordingly the ring structure is 
also rotating along +$\phi$. It is thus difficult to distinguish between the trMRN 
models with or without Hall effect, or to identify the polarity of the magnetic field, 
simply from the disc morphology and/or the direction of gas rotation. 
In fact, the three trMRN models also share a similar disc morphology and rotation 
direction with the anti-aligned model 2.4opt3\_AH$^-$O in \S~\ref{S.HallAnti}. 

The formation of RSDs and rings in the trMRN models is mainly facilitated by the 
enhanced ambipolar diffusivity that causes large outward drift of magnetic fields 
at the envelope scale \citep[see also][]{Zhao+2018a}. 
As shown in Fig.~\ref{Fig:2.4dust1kAH}, the effective velocity of the magnetic field 
lines in the radial direction $\varv_{{\rm B,eff},r}$ nearly vanishes between 
$\sim$20--500~AU along the pseudo-disc (equatorial plane) in all three models. 
While the radial drift of magnetic fields 
$\varv_{{\rm d},r}$ ($\equiv \varv_{{\rm B,eff},r}-\varv_r$) in the Hall-free model 
(2.4trMRN\_AO) is simply determined by the radial ambipolar drift that 
reaches $\sim$2--3~km~s$^{-1}$ in the inner tens of AU, the dominant radial drift 
component in the trMRN models with Hall effect remains the ambipolar drift 
$\varv_{{\rm AD},r}$ (indicated by $\varv_{{\rm B,eff},r}-\varv_{{\rm BH},r}$), 
which is around 2~km~s$^{-1}$ in the inner tens of AU. In comparison, the radial 
Hall drift $\varv_{{\rm H},r}$ (indicated by $\varv_{{\rm BH},r}-\varv_r$) is 
mostly vanishing along the pseudo-disc and only reaches a few 0.1~km~s$^{-1}$ in 
the innermost tens of AU. In any case, the net effect of ambipolar and Hall 
drift in the radial direction is to move the magnetic field outward relative to 
the infalling matter, for both the anti-aligned (.4trMRN\_AH$^-$O) and aligned 
(.4trMRN\_AH$^+$O) models. In particular, near $\sim$30~AU where $\varv_{{\rm H},r}$ 
is non-vanishing, the radial ambipolar and Hall drift velocities are both positive 
(i.e., pointing towards +$r$), indicating a cooperative effort of AD and Hall effect 
radially within that equatorial section. 

\begin{figure}
\includegraphics[width=1.0\columnwidth]{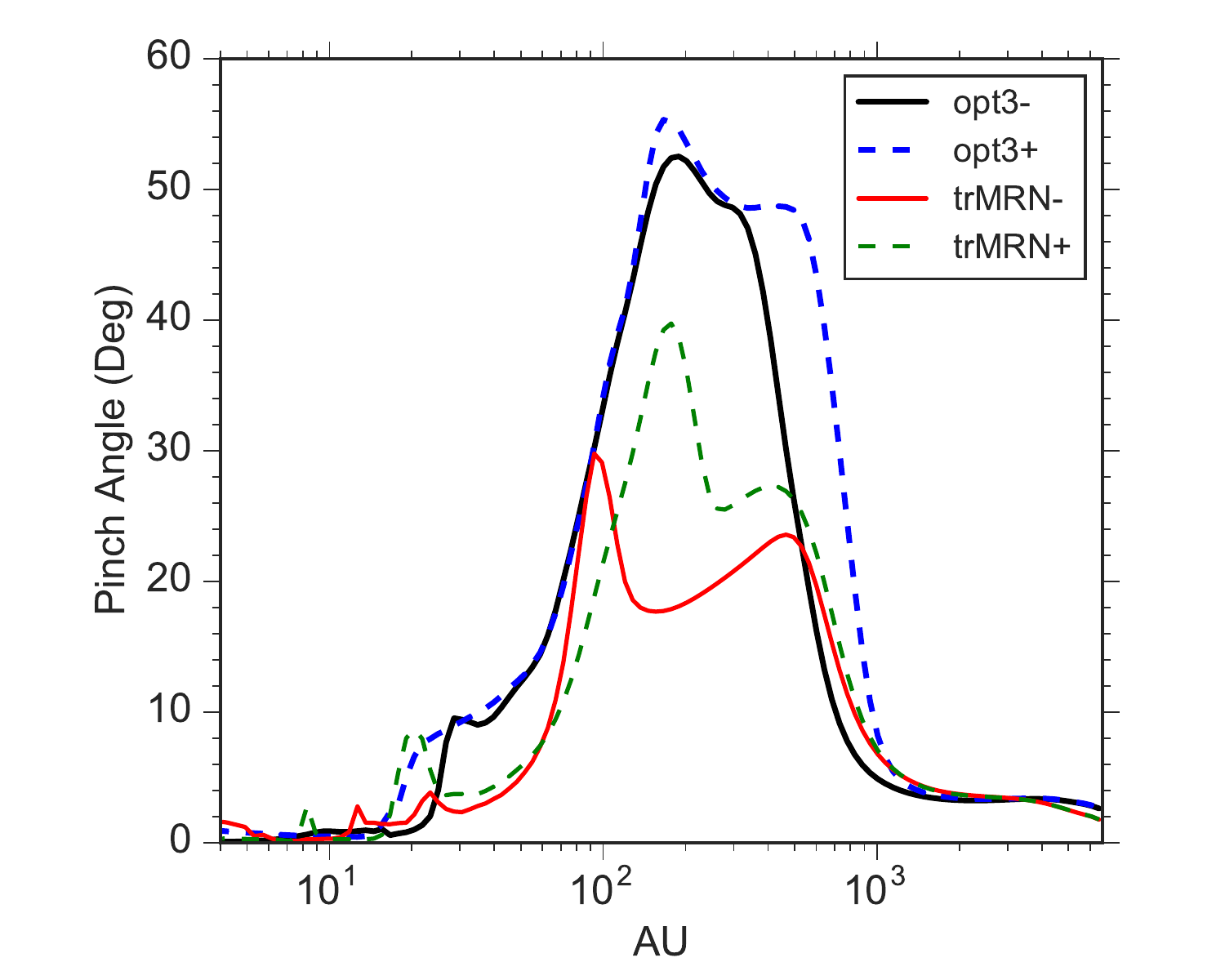}
\caption{Radial profile of pinch angle of magnetic field lines. A larger 
pinch angle indicates a more severe radial pinch of the magnetic field lines. 
Black solid, blue dashed, red solid, and green dashed lines correspond to 
model 2.4opt3\_AH$^-$O, 2.4opt3\_AH$^+$O, 2.4trMRN\_AH$^-$O, and 2.4trMRN\_AH$^+$O, 
respectively.}
\label{Fig:compAng}
\end{figure}

In the azimuthal direction, notable magnetic field drift only takes place in 
the inner 100~AU, a smaller region than the opt3 models. Although the azimuthal 
Hall drift $\varv_{{\rm H},\phi}$ (indicated by $\varv_{{\rm BH},\phi}-\varv_\phi$) 
is the main contributor to the total azimuthal drift $\varv_{{\rm d},\phi}$ 
($\equiv \varv_{{\rm B,eff},\phi}-\varv_\phi$), the magnitude of $\varv_{{\rm H},\phi}$ 
is also smaller than that in the opt3 models. Recall that the azimuthal Hall drift 
velocity $\varv_{{\rm H},\phi}$ is proportional to the degree of radial pinching of 
magnetic fields (Eq.~\ref{Eq:v_Hphi}), which is now relaxed by the efficient AD 
in the trMRN models. In Fig.~\ref{Fig:compAng}, we compare between the opt3 and trMRN 
models the pinch angle of magnetic field lines, which is defined as the angle between 
the vertical axis and the direction of the magnetic field just above and below the 
equatorial plane. The opt3 models show systematically larger pinch angles than 
the trMRN models by a factor of $\sim$2. The difference is the largest 
($\sim$20--30$^\circ$ difference) at the envelope scale (few 10$^2$~AU) where the 
enhanced AD in the trMRN models already becomes efficient in radially decoupling the 
magnetic field ($\varv_{{\rm B,eff},r} \rightarrow 0$) as compared to the much weaker 
AD in the opt3 models (Fig.~\ref{Fig:2.4opt3AH-} and~\ref{Fig:2.4opt3AH+}). 
At tens of AU scale, the factor of $\sim$2 difference in the pinch angle is still 
visible ($\sim$5$^\circ$ in the trMRN models vs. $\sim$10$^\circ$ in the 
opt3 models), which can also be recognized by comparing the magnetic field geometries 
among Fig.~\ref{Fig:2.4dust1kAH}, \ref{Fig:2.4opt3AH-}, and \ref{Fig:2.4opt3AH+}. 
Thus, as AD becomes efficient, both the degree of radial pinching of magnetic 
fields and the induced azimuthal Hall drift $\varv_{{\rm H},\phi}$ are reduced. 
For the same reason, the magnetic field strength near the ring-like structure 
is also weaker in the trMRN models than in the opt3 models, by a factor of $\sim$10, 
which causes the plasma-$\beta$ in the trMRN models to reach as high as 
$\sim$10$^4$--10$^5$. 

It is worth noting that within the inner tens of AU of the trMRN models with 
Hall effect, the azimuthal Hall and ambipolar drift operate counteractively, 
i.e., pointing towards opposite directions. For example, in the aligned model 
2.4trMRN\_AH$^+$O, the azimuthal Hall drift $\varv_{{\rm H},\phi}$ tends to 
increase the original bending of magnetic fields along +$\phi$ 
($\varv_{{\rm BH},\phi}-\varv_\phi>0$); however, the azimuthal ambipolar drift 
$\varv_{{\rm AD},\phi}$ sets to curb such a tendency by directing magnetic fields 
backward along -$\phi$ ($\varv_{{\rm B,eff},\phi}-\varv_{{\rm BH},\phi}<0$). 
In the anti-aligned model 2.4trMRN\_AH$^-$O, similar counteractive interplay 
between the azimuthal Hall and ambipolar drift is marginally visible near 
$\sim$30~AU, where magnetic fields become slightly bent towards -$\phi$ (so that 
the induced radial Hall drift is along +$r$). Note that, even with the enhanced 
ambipolar diffusivity, the 2.4trMRN\_AO model (no Hall effect) shows little 
azimuthal ambipolar drift ($\varv_{{\rm AD},\phi}\lesssim0.2$~km~s$^{-1}$), 
as compared to the large radial ambipolar drift $\varv_{{\rm AD},r}$, which also 
reflects that the bending of magnetic fields is preferentially stronger in the 
radial direction than in the azimuthal direction. 

\begin{figure}
\includegraphics[width=1.0\columnwidth]{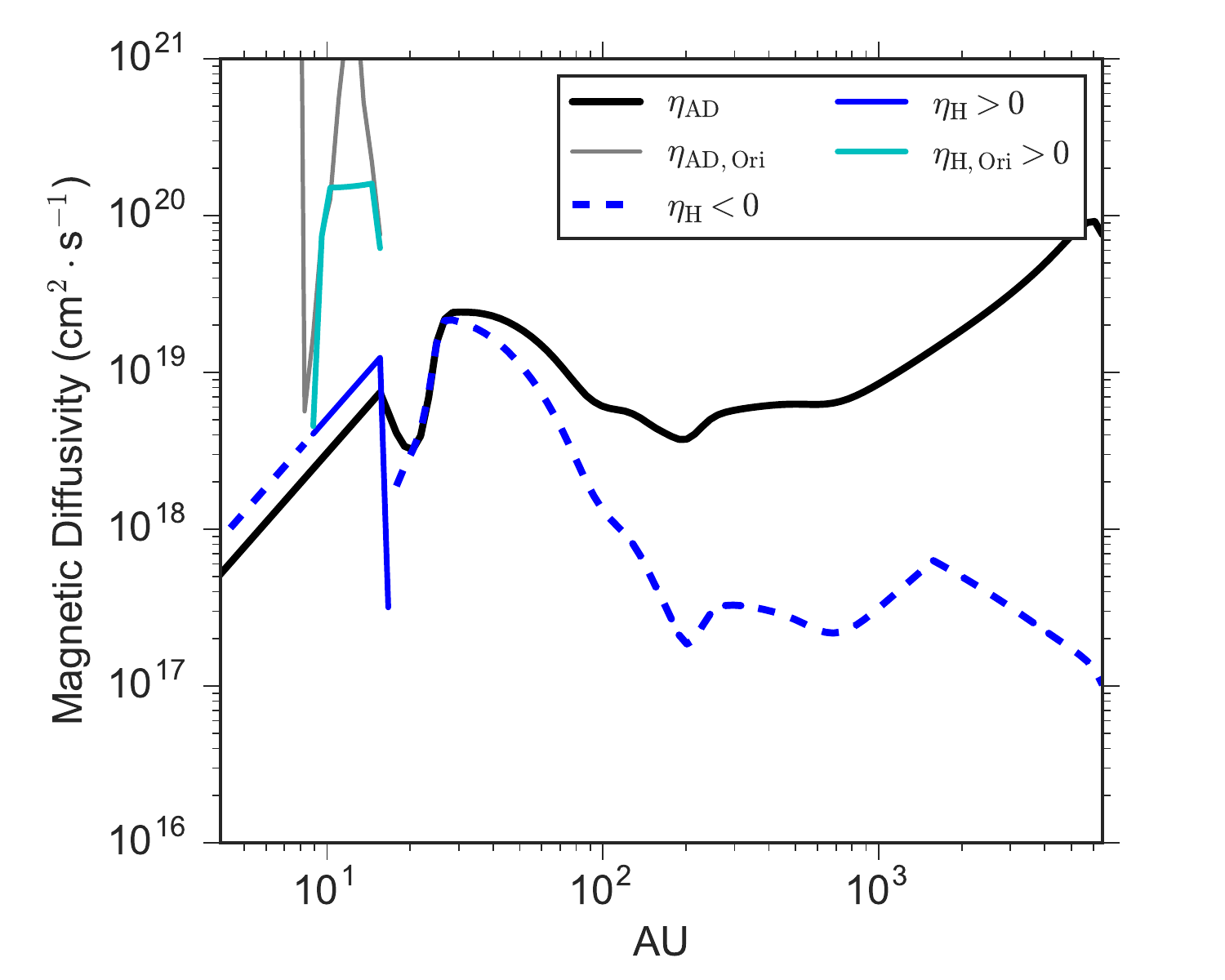}
\caption{Same as Fig.~\ref{Fig:2.4opt3eta}, but for the equatorial magnetic diffusivities 
of the 2.4trMRN\_AH$^+$O model at 134.674~kyr.}
\label{Fig:2.4trMRNeta}
\end{figure}

In summary, AD is dominating the magnetic field evolution and promoting disc 
formation in the trMRN models, in contrast to the opt3 models where Hall effect 
plays the dominant role instead. Such a role switch between AD and Hall effect 
is directly caused by the change of the size distribution $a_{\rm min}$ from 
0.03~$\mu$m to 0.1~$\mu$m. We show the magnetic diffusivities of the 2.4trMRN\_AH$^+$O 
model in Fig.~\ref{Fig:2.4trMRNeta}, in which the ambipolar diffusivity $\eta_{\rm AD}$ 
($\gtrsim$10$^{19}$~cm$^2$~s$^{-1}$), is $\sim$1--2 orders of magnitude larger than the 
Hall diffusivity $\eta_{\rm H}$ ($\sim$few 10$^{17}$ to few 10$^{18}$~cm$^2$~s$^{-1}$) 
throughout the envelope (outside the disc); $\eta_{\rm H}$ only starts to catch up with 
$\eta_{\rm AD}$ within the inner tens of AU. Recall that in the opt3 models 
(Fig.~\ref{Fig:2.4opt3eta}), $\eta_{\rm H}$ is instead dominating over $\eta_{\rm AD}$ 
in the inner envelope. Moreover, $\eta_{\rm H}$ of the trMRN models only becomes 
comparable to that of the opt3 models within the inner 100~AU scale, where the Hall 
drift velocities become notable in Fig.~\ref{Fig:2.4dust1kAH} of the trMRN models.

Finally, as a result of the strong AD and weak Hall effect in the trMRN models, 
there is no clear signature of counter-rotating motion in the collapsing envelope, 
in contrast to the well-established counter-rotating regions in both the aligned 
and anti-aligned opt3 models (see \S~\ref{S.Envelope}).

\subsection{High Cosmic-Ray Ionization Rate: Reduced Ion-neutral Drift \& Small Disc}
\label{S.highCR}

\begin{figure}
\hspace*{-0.20in}\includegraphics[width=1.25\columnwidth]{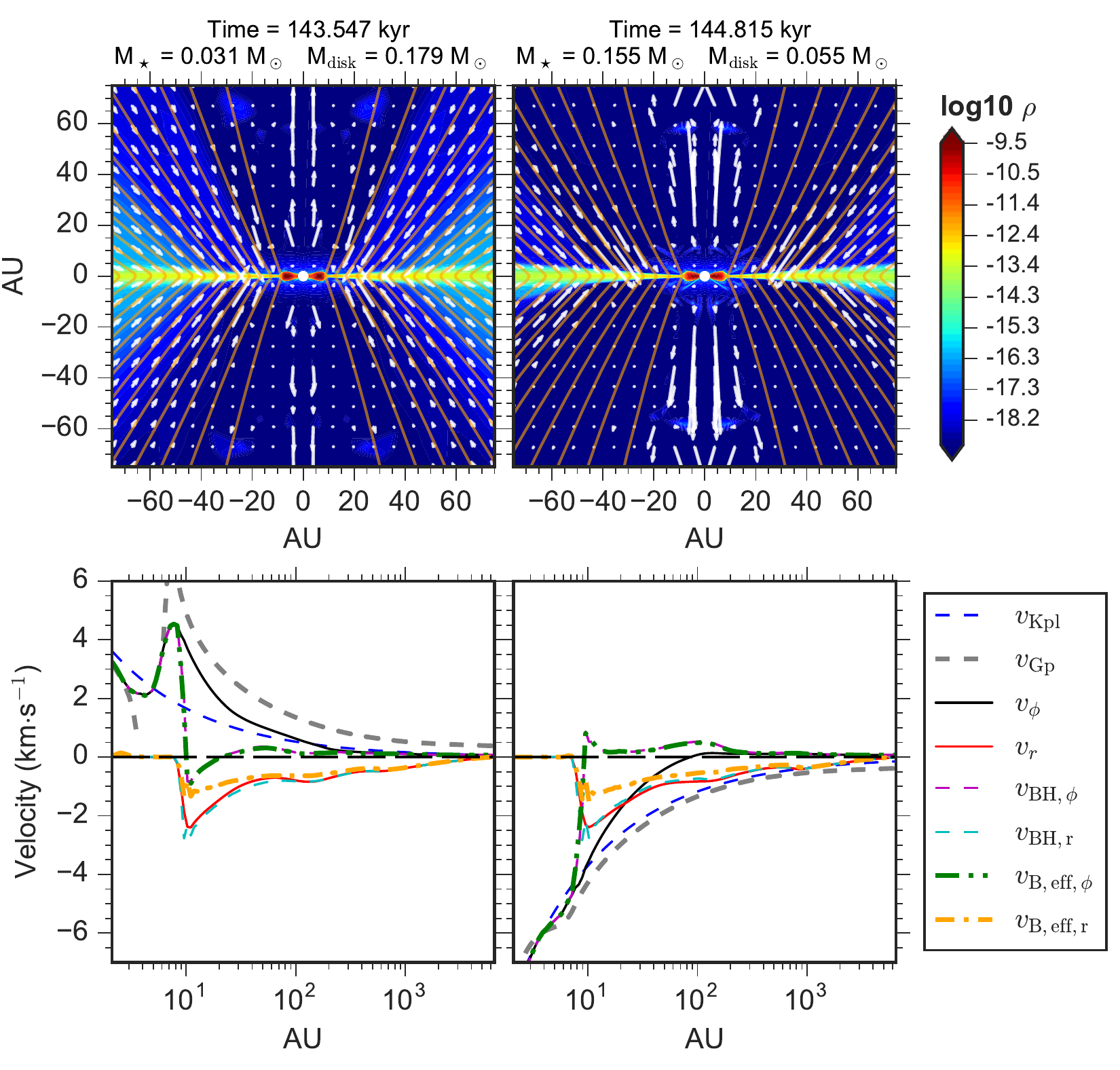}
\begin{tabular}{lcc}
& \hspace*{0.12in} (a) Anti-aligned: $\bmath{\Omega \cdot B}<0$ & \hspace*{0.14in} (b) Aligned: $\bmath{\Omega \cdot B}>0$\\
\end{tabular}
\caption{Mass density distribution (top) and velocity profile along the equator 
(bottom) for model 2.4CR10opt3\_AH$^-$O (left panels) and model 2.4CR10opt3\_AH$^+$O 
(right panels). White arrows and orange lines in the top panel are the velocity field 
vectors and magnetic field lines, respectively.}
\label{Fig:2.4CR10}
\end{figure}

We now explore the case of higher CR ionization rate, i.e., 
$\zeta_0^{\rm H_2}=10^{-16}$~s$^{-1}$, with which the magnetic diffusivities are 
overall lower by a factor of $\sim$2--3 than with the canonical $10^{-17}$~s$^{-1}$ 
($\propto \sqrt{\zeta_0^{\rm H_2}}$; \ct{UmebayashiNakano1990}) throughout the 
collapsing envelope. In either the aligned or anti-aligned case (Fig.~\ref{Fig:2.4CR10}), 
a small compact RSD is formed, with radius remained around $\sim$10~AU till the 
end of the simulation ($\sim$20~kyr after the first core formation). Despite 
the reduced diffusivity, the azimuthal Hall drift velocity can still efficiently 
regulate the magnetic field in the azimuthal direction, causing the gas rotation 
in the inner envelope to either spin-up or down depending on the polarity of the 
magnetic field. Similar to the opt3 models with a canonical $\zeta_0^{\rm H_2}$ 
(\S~\ref{S.HallRegime}; canonical-opt3 models for short), the resulting disc can either 
rotate along +$\phi$ in the anti-aligned case (2.4CR10opt3\_AH$^-$O) or along -$\phi$ in 
the aligned case (2.4CR10opt3\_AH$^-$O). 

As shown in Fig.~\ref{Fig:2.4CR10}, the total azimuthal drift 
is primarily determined by the azimuthal Hall drift, while the azimuthal 
ambipolar drift is vanishing. The magnitude of the azimuthal Hall drift velocity 
$\varv_{{\rm H},\phi}$ is somewhat smaller (by a factor of $\sim$2) in comparison 
to the opt3 models with a canonical $\zeta_0^{\rm H_2}$ (Fig~\ref{Fig:2.4opt3AH-} 
and \ref{Fig:2.4opt3AH+}). In the radial direction, the total radial drift is 
dominated by the radial ambipolar drift, whereas the radial Hall drift is negligible. 
The radial ambipolar drift velocity $\varv_{{\rm AD},r}$ is also somewhat smaller 
than that of the canonical-opt3 models; however, the radial ion-neutral decoupling 
already starts in the low density envelope from $\sim$800~AU --- similar scale as 
in the canonical-opt3 models. Although the radial Hall drift is barely non-vanishing 
within the inner $\lesssim$100~AU scale, its direction is consistent with that 
demonstrated in the canonical-opt3 models in \S~\ref{S.HallRegime}, i.e., $\varv_{{\rm H},r}$ 
in the inner envelope points radially inward in the anti-aligned case and radially 
outward in the aligned case. Nevertheless, the total radial drift $\varv_{{\rm d},r}$ 
dominated by AD is still moving the magnetic field radially outward relative to the 
infalling matter, but in a reduced speed compared with the opt3 models with a canonical 
$\zeta_0^{\rm H_2}$. 

\begin{figure}
\includegraphics[width=1.0\columnwidth]{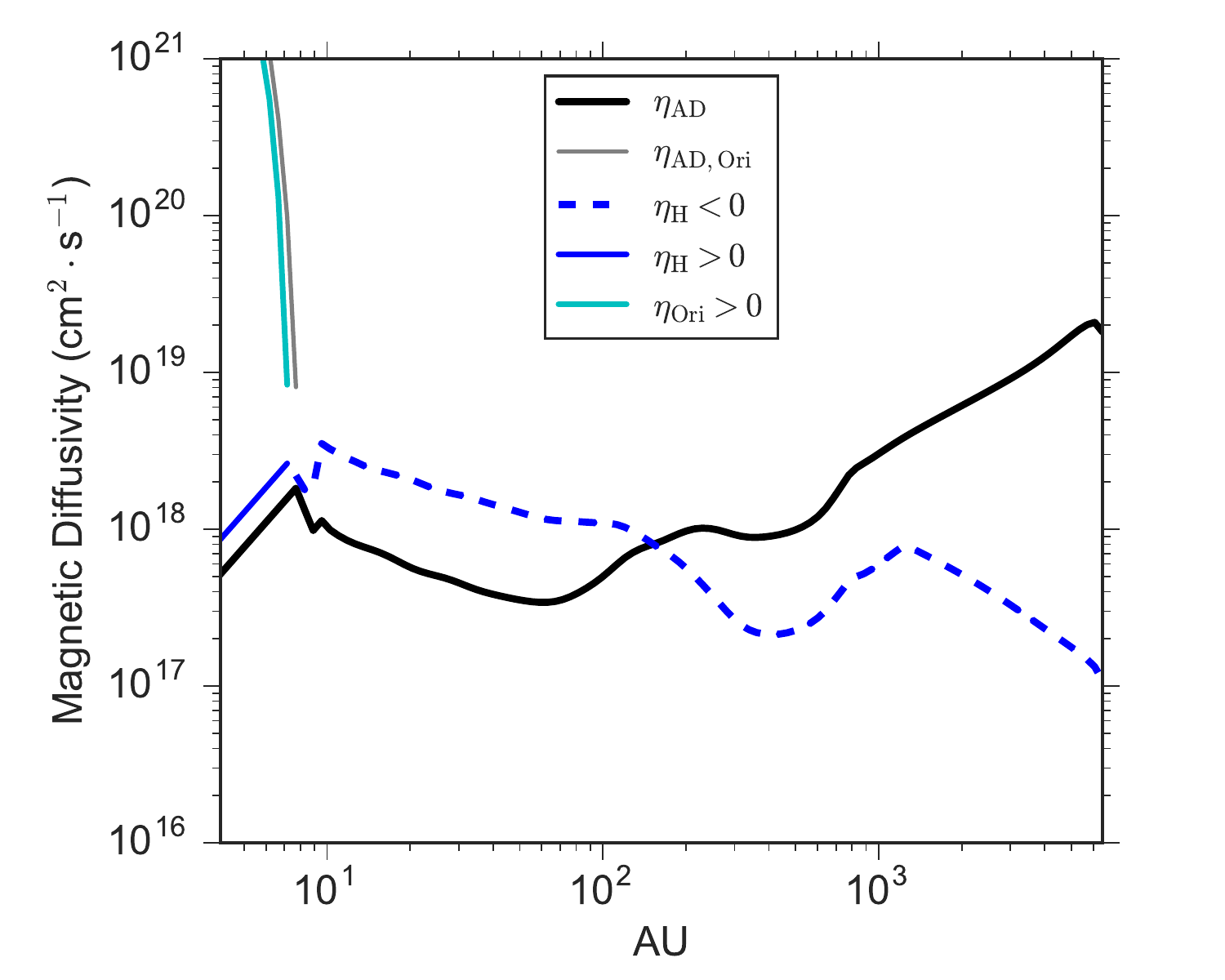}
\caption{Same as Fig.~\ref{Fig:2.4opt3eta}, but for equatorial magnetic diffusivities 
of the 2.4CR10opt3\_AH$^+$O model at 144.815~kyr.}
\label{Fig:2.4CR10eta}
\end{figure}

In terms of disc formation, these opt3 models with a higher CR ionization rate 
($\zeta_0^{\rm H_2}=10^{-16}$~s$^{-1}$) lie in between the MRN models 
(\S~\ref{S.MRN}) and the opt3 models (\S~\ref{S.HallRegime}) with the canonical CR 
ionization rate. This is caused by the variations in microphysics. 
In comparison to the MRN models (Fig.~\ref{Fig:2.4MRNeta}), these opt3 models with 
a high CR ionization rate (Fig.~\ref{Fig:2.4CR10eta}) still have a twice larger 
$\eta_{\rm H}$ within the inner $\lesssim$100~AU (a few 10$^{18}$~cm$^2$~s$^{-1}$), 
and a $\sim$10 times larger $\eta_{\rm AD}$ in the envelope scale beyond $\gtrsim$100~AU. 
The radial ambipolar drift throughout the infalling envelope along with the azimuthal 
Hall drift in the innermost envelope together enable the formation of the small 
compact RSDs in these high CR models. We have explored models with even higher 
$\zeta_0^{\rm H_2}$, and find that when $\zeta_0^{\rm H_2}$ is above 
2--3$\times$10$^{-16}$~s$^{-1}$, $\eta_{\rm AD}$ and $\eta_{\rm H}$ becomes 
comparable to that of the MRN models, so that disc formation is strongly suppressed 
in such axisymmetric set-ups. 
Note that the conditions of $\zeta_0^{\rm H_2}$ for disc formation is slightly 
more stringent if Hall effect is excluded, with a threshold of a few 10$^{-17}$~s$^{-1}$ 
derived in \citet[][see also \ct{Kuffmeier+2020}]{Zhao+2016}. 

\subsection{Impact of Initial Conditions \& Grain Size Limit}
\label{S.ICImpact}

Previous sections focus on the representative models in the strong field case 
($\lambda \sim 2.4$); in this section, we summarize results of the other models 
in Table~\ref{Tab:model1}--\ref{Tab:model3}, and demonstrate how AD and Hall effect 
affect disc formation and morphology in different parameter space.

As shown in \PaperI~ \citep[see also][]{WursterBate2019}, decreasing the initial 
magnetic field strength weakens Hall effect in general. For the opt3 models 
($a_{\rm min}=0.03$~$\mu$m) in which Hall drift is the most efficient, 
disc morphology in the anti-aligned case is relatively unaffected by the change 
of magnetic field strength; the anti-aligned opt3 models mostly form 20--30~AU 
ring-like structures (spirals/multiples in 3D that becomes more gravitationally 
unstable as the core magnetization weakens; \ct{Zhao+2018a}). 
However, for the aligned opt3 models, weakening the initial magnetic field 
strength indeed hinders the formation of the counter-rotating disc.\footnote{For 
the aligned model 4.8opt3\_A$^{+}$H, a normally rotating disc of $\sim$15~AU 
forms initially but shrinks over time towards the central stellar object. 
It is possible that the star itself (and materials close to the star) can spin in 
the opposite direction of the counter-rotating disc, but the same direction as the 
bulk envelope material.}
For example, the radius of the counter-rotating disc 
is only $\sim$20~AU in the 4.8opt3\_AH$^+$O model, compared with the $\sim$30~AU 
counter-rotating disc in the 2.4opt3\_AH$^+$O model. In the very weak field 
($\lambda\sim9.6$) model 9.6opt3\_AH$^+$O, no clear counter-rotating motion develops 
in the inner envelope and only a ring-like structure forms, which rotates along the 
direction of initial core rotation (+$\phi$); it is thus difficult to determine 
the magnetic field polarity from the disc morphology in the very weak field case. 
Basically, in weaker field models, the azimuthal Hall drift becomes less efficient, 
and the tension force directed towards -$\phi$ in the aligned cases also becomes weaker, 
or even fails to brake the original gas rotation along +$\phi$ in the very weak 
field case ($\lambda\sim9.6$). The main role of Hall effect in the 9.6opt3\_AH$^+$O 
model is instead to drift the magnetic field radially outward, similar to the weak 
field model (without AD) presented in \PaperI. 

For the trMRN models in which AD is the most efficient, weakening the initial magnetic 
field strength results in somewhat larger RSDs or ring-like structures, consistent 
with the parameter study of \citet{Zhao+2016,Zhao+2018a} without Hall effect. 
Again, the disc morphology across the trMRN models is insensitive to Hall effect and 
hence the magnetic field polarity, as the poloidal magnetic fields are efficiently 
drifted radially outward by the enhanced AD (see \S~\ref{S.ADRegime}). 

For the MRN models and the models with very high-CR ionization rate 
($\zeta_0^{\rm H_2}=5 \times 10^{-16}$~s$^{-1}$) where disc formation is 
suppressed, weakening the initial magnetic field strength increases the radius of 
the initial disc structure formed shortly after the first core. However, with the 
low level of magnetic diffusivities, the initial disc structures in the MRN models 
and the high CR models typically shrink in radius over time, as collapse drags more 
magnetic flux into the central disc forming region 
\citep[see also][]{Zhao+2016,Zhao+2018a,Lam+2019}. 
It is only in models with $\zeta_0^{\rm H_2}=10^{-16}$~s$^{-1}$ that a small 
compact disc of $\sim$10 radius survives, for both strong and weak magnetic field. 

We also explore the collapse of initially non-rotating cores for the grain 
size distributions of opt3 and trMRN. Sizable RSDs of $\sim$30~AU radius is able to 
form in the non-rotating opt3 models (both strong and weak field) due to the large 
Hall diffusivity. The rotational motion in the inner envelope and hence the disc is 
essentially generated by Hall effect, which causes the radially pinched magnetic field 
lines to become bended azimuthally along the direction of the azimuthal Hall drift 
(+$\phi$). The resulting tension force then spins up the gas rotation in the inner 
envelope towards the opposite direction (-$\phi$) of the azimuthal Hall drift. 
In contrast, in the non-rotating trMRN models (either strong or weak field) 
where AD dominates the collapse, the weak Hall effect only slightly spins up the gas 
rotation along -$\phi$, but not enough to form any rotationally supported structure. 

Finally, we find that the lower limit of $a_{\rm min}$ above which disc formation 
by non-ideal MHD effects becomes possible, is around $\lesssim$10~nm, similar to 
that discussed in \PaperI~ where AD is ignored. Note that such a limit is less 
stringent than that derived in the Hall-free study of \citet{Zhao+2016}, where 
the lower limit of $a_{\rm min}$ was a few times larger. For the min1 models in 
Table~\ref{Tab:model1}--\ref{Tab:model2} with $a_{\rm min}$ set to 10~nm, 
the ambipolar diffusivity is not much different from the MRN models, and efficient 
radial drift by AD only occurs along the pseudo-disc within the inner $\lesssim$100~AU 
scale. However, the Hall diffusivity is already enhanced by $\sim$1 order of magnitude 
relative to the standard MRN size distributions at densities above 
$\gtrsim$10$^9$~cm$^{-3}$ (corresponding to the inner $\lesssim$200~AU pseudo-disc 
region; see also \ct{Dzyurkevich+2017,Zhao+2018b}). 
Therefore, both AD and Hall effect are efficient within $\lesssim$100~AU scale, which 
enables formation of RSDs in the min1 models. Nonetheless, the requirements on 
microphysics for disc formation, particularly $a_{\rm min}$ and $\zeta_0^{\rm H_2}$ 
(\S~\ref{S.highCR}), are somewhat relaxed by the inclusion of Hall effect, in 
comparison to studies focused only on AD.

\section{Discussion}
\label{Chap.Discuss}

\subsection{Comparison with Hall-only Models}

We have shown in \PaperI~ that, in the absence of AD, Hall effect can only produce 
small $\lesssim$10--20~AU discs, regardless of the magnetic field polarity. 
The small disc radii are caused by the development of the so-called ``RSHCS'' 
(rotationally supported Hall current sheet) surrounding the inner disc during the 
main accretion phase. Within the RSHCS, highly-pinched poloidal magnetic fields are 
moving radially inward relative to matter, flattening the RSHCS region into a 
thin sheet-like structure. 

In this study, the presence of AD largely suppresses the development of RSHCS,\footnote{
Transient RSHCS-like features do occasionally appear in this study, 
but are only limited to narrow regions (width $\lesssim$1~AU) near the 
disc edge, where the total radial drift points briefly inward. Such transient 
features is unlikely to affect the disc formation, yet they may not be numerically 
resolved with the current set-up.}
as the radial ambipolar drift efficiently slows down the inward drift of 
magnetic fields by Hall effect at tens of AU scale, so that the total radial 
drift is mostly pointing outward (Fig.~\ref{Fig:2.4opt3AH-} 
and~\ref{Fig:2.4opt3AH+}). The RSDs formed in this study are generally larger 
in radius than in \PaperI, and would likely develop extended spiral structures 
or become multiple systems in full 3D set-ups \citep{Zhao+2018a,WursterBate2019}. 
Therefore, persistent outward diffusion of magnetic field in the radial direction 
remains crucial for disc formation and growth. 

Another point worth noting is that, the processes of protostellar collapse 
and disc formation between the opt3 and trMRN models in \PaperI~ 
(without AD) behave similarly, with the opt3 models forming slightly larger 
disc sizes than the trMRN models (see Table 1 \& 2 in \PaperI) due to the 
differences in $\eta_{\rm H}$. However, as we have shown in this study (with AD), 
changing the minimum grain size from 0.03~$\mu$m (opt3) to 0.1~$\mu$m (trMRN) 
causes the collapse to switch from Hall-dominated to AD-dominated; the morphologies 
and kinematics of the envelope and disc show clear differences between 
the two types of collapse.

\begin{figure*}
\includegraphics[width=\textwidth]{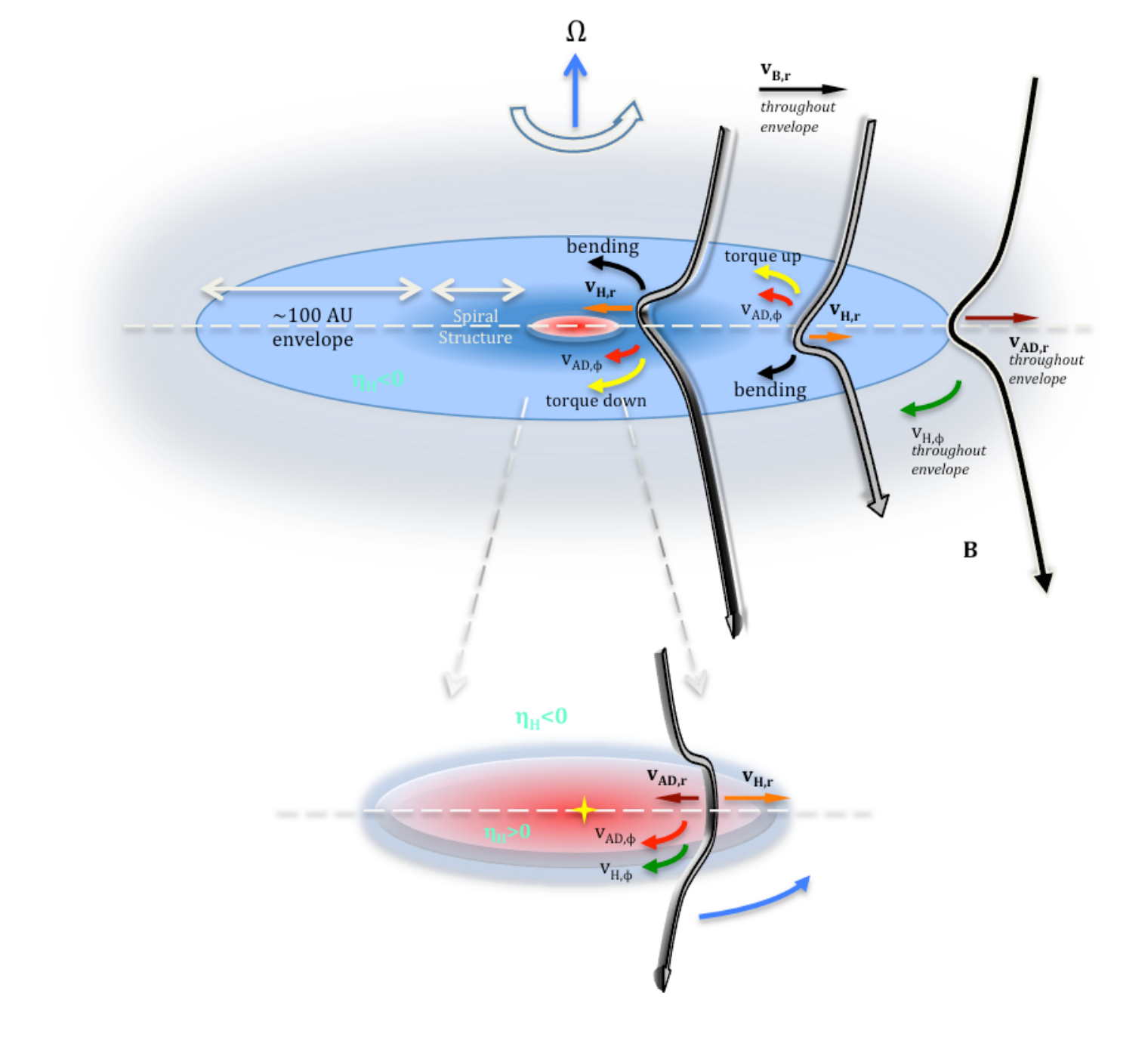}
\caption{Sketch of magnetic field morphologies in the inner envelope (top) and 
in the disc (bottom) for the anti-aligned case ($\bmath{\Omega \cdot B}<0$) in the 
Hall dominated collapse. Note that $\eta_{\rm H}$ changes sign across the disc boundary, 
and that the actual value of $\varv_{{\rm AD},\phi}$ in the top panel is vanishing 
(see Fig.~\ref{Fig:2.4opt3AH-}).}
\label{Fig:Hall-B}
\end{figure*}

\begin{figure*}
\includegraphics[width=\textwidth]{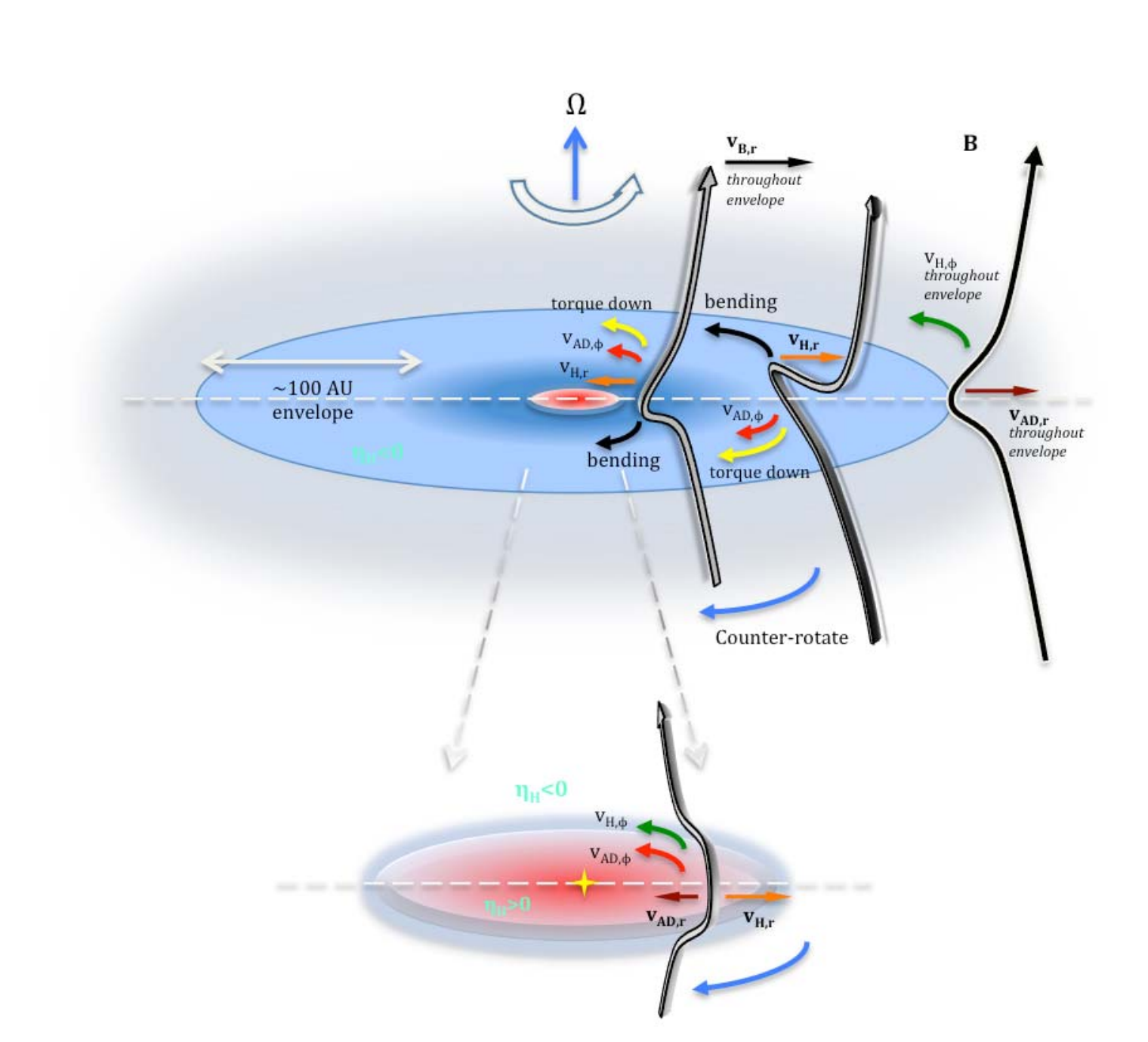}
\caption{Same as Fig.~\ref{Fig:Hall-B}, but for the aligned case ($\bmath{\Omega \cdot B}>0$) 
in the Hall dominated collapse. Note that in the top panel, the bending of the leftmost 
field line towards -$\phi$ tends to torque down the counter-rotation.}
\label{Fig:Hall+B}
\end{figure*}

Similar to \PaperI, we illustrate the general configuration of the magnetic 
field under the complex regulation of ambipolar and Hall drift, 
for the Hall dominated collapse (the illustration of an AD dominated collapse 
is much simpler). In Fig.~\ref{Fig:Hall-B} and 
Fig.~\ref{Fig:Hall+B}, the directions of ambipolar and Hall drift in different 
sections of the inner envelope and disc follow the same basic principles derived 
in \S~\ref{S.ADHallDrift} (see Fig.~\ref{Fig:sketch}). The combined effort of 
AD and Hall effect is to drift the magnetic field against the primary direction 
of field bending by the dominant dynamical process, i.e., gravitational collapse in 
the envelope or Keplerian rotation in the disc. Note that due to the severe 
azimuthal bending of magnetic fields (${\partial B_\phi \over \partial z}$) inside 
the disc, the radial Hall drift usually dominates the radial ambipolar drift therein. 

It is also clear from such illustrations that the disc formed from a Hall dominated 
collapse can only have one particular magnetic field polarity ($\bmath{\Omega \cdot B}<0$; 
see also discussions in \PaperI). Accordingly, the magnetic field along the disc 
mid-plane tends to drift radially outward by Hall effect \citep{BaiStone2017}. 
However, this would not be the case for an AD dominated collapse or a weakly 
magnetized core, since the eventual disc rotation is little affected by Hall effect 
during the collapse, and both $\bmath{\Omega \cdot B}<0$ and $\bmath{\Omega \cdot B}>0$ 
become possible. 
Therefore, the magnetic field polarity as well as the direction of field diffusion 
in protoplanetary discs are closely related to how discs are formed from the 
protostellar collapse and what microphysics are considered in the protostellar envelope.
Finally, across the disc boundary (either radially or vertically), $\eta_{\rm H}$ 
changes sign as Hall drift switches from being dominated by the drift of 
positively-charged species (e.g., ions) relative to negatively-charged species 
(e.g., negatively-charged grains) to being dominated by the usual electron-ion drift 
(\PaperI; see also \ct{XuBai2016,Lesur+2020}), which could potentially affect the 
development of the so-called ``Hall-shear'' instability \citep{Kunz2008} across 
the disc surface \citep{BaiStone2017}.

\subsection{Disc to Stellar Mass Ratio}

Across the Hall dominated models with $a_{\rm min}=0.03~\mu$m, not only the direction 
of disc rotation is bimodal, the disc to stellar mass ratio also bifurcates: 
the ratio is much larger than unity in the anti-aligned case ($\bmath{\Omega \cdot B}<0$), 
and is lower or close to unity in the aligned case ($\bmath{\Omega \cdot B}>0$). 
The difference is mainly caused by the amount of mass accumulated in the central 
stellar object at early times shortly after the first core formation. 
Although the 2D axisymmetric set-up generally hinders the stellar mass accretion, 
a considerable amount of mass (0.1--0.2~$M_{\sun}$) is able to accrete onto the 
central stellar object in the aligned opt3 models (\S~\ref{S.HallAlign}) during 
the initial disc suppression phase.
In contrast, the stellar mass remains relatively small in the anti-aligned opt3 
models and the RSDs are mostly self-gravitating. 

However, we expect such a bifurcation of disc to stellar mass ratio to be somewhat 
alleviated in full 3D simulations, as asymmetric structures such as spirals promote 
the mass accretion into the central stellar vicinity for both the anti-aligned 
and aligned cases. In fact, how mass is funneled from discs to protostars remains 
a debated topic \citep{Takasao+2018}. With a limited resolution in protostellar 
collapse simulations, artificial sinks are usually introduced to model the central 
stellar object. In this study, we consider mass falling into the inner boundary 
($r_{\rm in}=2$~AU) as being accreted by the central stellar object. It is obvious 
that the larger the inner boundary, the easier it is for the central stellar object 
to grow in mass (e.g., \ct{Krasnopolsky+2011} and \ct{Li+2011} adopted 
$r_{\rm in} \sim 6.67$~AU). As a result, the disc to stellar mass ratio is also 
dependent on the size of the inner boundary. Similar phenomenon is often discussed 
in studies using sink particles to model the central stellar object 
\citep{Machida+2014,Hennebelle+2020}, in which typical parameters of the sink 
treatment including accretion radius and threshold density, can have a substantial 
impact on both disc formation \citep{Machida+2014} and disc to stellar mass 
ratio \citep{Hennebelle+2020}. Essentially, decreasing the threshold density and/or 
increasing the accretion radius facilitate the mass growth of the central stellar 
object, but suppress the disc to assemble mass, especially in the ideal MHD limit. 

Nevertheless, it would be important to explore in 3D the relation between the 
disc to stellar mass ratio and the magnetic field polarity in a Hall dominated 
collapse, as well as how much the inner boundary or sink treatment can affect 
such a relation. Future studies along this line could offer critical guidance 
for observations to constrain the masses of Class 0/I protostellar discs 
\citep{Manara+2018,BalleringEisner2019}.

\subsection{Outflow Morphology}

\begin{figure*}
\centerline{\includegraphics[width=1.1\textwidth]{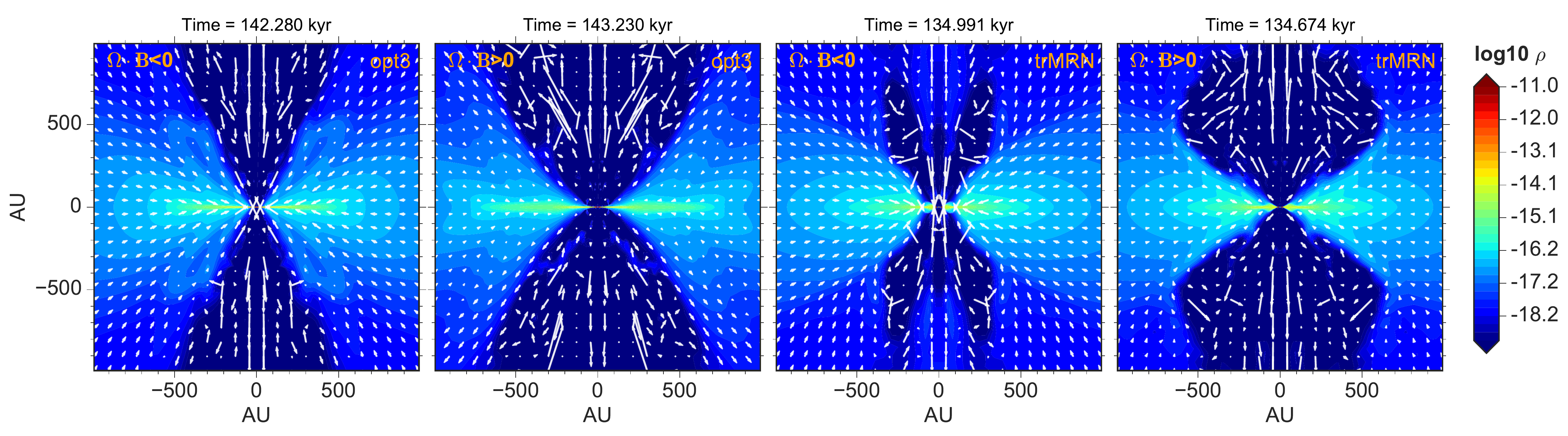}}
\caption{Bipolar outflows at 1000~AU scale for model 2.4opt3\_AH$^-$O at $t=142.280$~kyr 
(1st panel from left), 2.4opt3\_AH$^+$O at $t=143.230$~kyr (2nd panel), 2.4trMRN\_AH$^-$O 
at $t=134.991$~kry (3rd panel), and 2.4trMRN\_AH$^+$O at $t=134.674$~kyr (4th panel), 
respectively. Velocity field vectors are shown in white arrows. The total mass of 
star and disc is similar among the four frames, implying a similar evolution stage.}
\label{Fig:outflows}
\end{figure*}

As microphysics modulates the non-ideal MHD effects, not only the collapse process 
and disc formation are profoundly affected, but the accompanying bipolar outflows 
also manifest diverse morphologies and properties. 
As shown in Fig.~\ref{Fig:outflows}, the aligned cases ($\bmath{\Omega \cdot B}>0$) 
preferentially show wider outflow cavities than the anti-aligned cases 
($\bmath{\Omega \cdot B}<0$), with either opt3 or trMRN grains. 
This is an outcome of the regulation of $B_\phi$ by the azimuthal Hall drift, 
which tends to enhance $B_\phi$ towards the direction of the initial rotation 
+$\phi$ in the aligned cases, but to weaken $B_\phi$ with $\varv_{{\rm H},\phi}$ 
pointing towards -$\phi$ in the anti-aligned cases 
(see Fig.~\ref{Fig:2.4opt3AH-}, \ref{Fig:2.4opt3AH+} and ~\ref{Fig:2.4dust1kAH}). 
Such an enhancement and weakening of the toroidal magnetic field by Hall effect 
then determines the strength and open-angle of the centrifugal driven outflows 
\citep{BlandfordPayne1982,PelletierPudritz1992,Tomisaka2002,Seifried+2012}. 
Indeed, the outflow velocity at few 100~AU scale above the disc plane is only 
$\sim$1--2~km~s$^{-1}$ in the anti-aligned model 2.4opt3\_AH$^-$O, which is a 
factor of 2--3 slower than in the aligned model 2.4opt3\_AH$^+$O 
($\sim$4--6~km~s$^{-1}$ at the same scale). Furthermore, because the disc 
in the aligned model is counter-rotating, the outflow region near the bipolar 
axis is also counter-rotating with respect to the bulk envelope (see 
Fig.~\ref{Fig:envelopVph}); this can be an observable feature \citep{Takakuwa+2018} 
for identifying the polarity of the magnetic field in the Hall dominated collapse. 

Furthermore, switching the grain size distribution from opt3 to trMRN generally 
reduces the outflow velocity by a factor of $\sim$2 for both polarities of the 
magnetic field. The strong AD operating in the trMRN models decreases the amount 
of magnetic flux dragged into the inner envelope 
\citep[hence reducing the magnetic field strength;][]{Zhao+2018a}, and also lowers 
the degree of radial pinching of poloidal magnetic fields (see Fig.~\ref{Fig:compAng}); 
both factors tend to suppress the launching of centrifugal driven outflows 
\citep[see also discussions in][]{Tsukamoto+2020,Marchand+2020}. 
In Fig.~\ref{Fig:outflows}, the outflow cavities in the trMRN models extend to a 
lower height vertically than in the opt3 models. Note that the trMRN models mostly 
form ring-like structures (spirals/multiples in 3D), which may also disfavour strong 
outflows in comparison to a compact disc structure.

\subsection{Comparison with Other Studies}

The importance of microphysics in disc formation, including grain size distribution 
and CR ionization rate, has only been explored by different groups since recently 
\citep{Padovani+2014,Zhao+2016,Tsukamoto+2020,Marchand+2020}, as earlier work 
suggested that an enhanced resistivity is necessary to enable the formation of 
sizable RSDs \citep{Shu+2006,Krasnopolsky+2010}. However, difference in the 
chemistry networks and cloud initial conditions can often lead to different 
conclusions. 

\citet{Tsukamoto+2020} showed that tens of AU RSDs can still form with the standard 
MRN size distribution, which is mainly due to their choice of a centrally concentrated 
density profile (Bonnor-Ebert sphere) combined with a uniform magnetic field. Such 
an initial set-up is shown by previous studies \citep{Machida+2014,Lam+2019} 
to facilitate disc formation even with the fiducial level of ambipolar diffusivity 
from \citet{Shu1991}, whose magnitude is comparable to that derived using the 
standard MRN size distribution at the envelope density (10$^6$--10$^9$~cm$^{-3}$; 
\ct{Zhao+2018b}). As pointed out by \citet{Machida+2014}, the mass-to-flux ratio 
near the cloud centre is higher in the Bonner-Ebert sphere set-up than in the 
uniform sphere set-up, by a factor of $\sim$2--3, even if the global mass-to-flux 
remains the same for the two set-ups. Ideally, a more self-consistent distribution 
of the magnetic field strength and geometry (e.g., ``hourglass-shaped'') should be 
initialized for clouds with a centrally concentrated density profile; otherwise, 
initializing a uniform magnetic field would have assumed that mass has concentrated 
in the cloud centre without dragging in the magnetic field. 

Furthermore, the ambipolar diffusivities obtained from the ionization chemistry of 
\citet{Tsukamoto+2020} is mostly lower than Shu's fiducial $\eta_{\rm AD}$ level by 
1--2 orders of magnitude, the opposite to the results of other studies 
\citep[e.g.,][]{Dzyurkevich+2017,Zhao+2018b}. In fact, disc formation has been 
shown to be strongly suppressed with an $\eta_{\rm AD}$ profile below the Shu's 
fiducial level \citep{MellonLi2009,Li+2011,Lam+2019}. Hence, it is important to 
ensure that the ionization chemistry converges among different groups. 
Note that it is also unclear in \citet{Tsukamoto+2020} how the magnetic flux is 
treated when matter is accreted onto the sink particle, the lack of ``DEMS'' 
\citep[Decoupling-Enabled Magnetic Structures;][]{Zhao+2011,Krasnopolsky+2012,MachidaBasu2020,Hennebelle+2020} 
across their models may suggest a deletion of the magnetic flux associated with the 
accreted matter \citep[similar to][]{Wurster+2017}, which further helps 
RSDs to form and survive even with a relatively low magnetic diffusivity. 

The series of non-ideal MHD core collapse simulations using RAMSES 
\citep{Masson+2016,Hennebelle+2020,Marchand+2020} mostly adopt the ionization 
chemistry from \citet{Marchand+2016}, in which a modified MRN size distribution 
($a_{\rm min} = 0.0181~\mu$m and $a_{\rm max} = 0.9049~\mu$m) is used to compute 
the magnetic diffusivities. As pointed out by \citet{Zhao+2016}, such a modified MRN 
distribution already over-performs (in terms of $\eta_{\rm AD}$) the standard MRN 
distribution as well as the Shu's fiducial level by up to a factor of 10 at the 
envelope density. Tens of AU RSDs are expected to form under such an enhanced 
diffusivity. The analytical formula derived in \citet{Hennebelle+2016} adopts 
a relatively high $\eta_{\rm AD}$ of $7.16 \times 10^{18}$~cm$^2$~s$^{-1}$. 
In addition, \citet{Hennebelle+2020} set an initial misalignment angle 
of 30$^\circ$ between the magnetic field and rotation, which also makes disc formation 
easier than the axisymmetric set-ups \citep{Joos+2012,Li+2013}. 

In comparison to our previous work \citep{Zhao+2016,Zhao+2018a}, the inclusion of 
Hall effect relaxes the requirements on microphysics for disc formation 
(\S~\ref{S.highCR} and \S~\ref{S.ICImpact}). In particular, 
(1) the lower limit of $a_{\rm min}$ reduces from a few times 10~nm to 
just $\lesssim$10~nm, i.e., 
truncating off the tiny grains (e.g., PAHs) at the lower end of the MRN size 
distribution is already enough for non-ideal MHD effects to promote disc formation; 
(2) the upper limit of $\zeta_0^{\rm H_2}$ increases to 2--3$\times$10$^{-16}$~s$^{-1}$, 
a factor of 10 larger than the limit derived in our Hall-free study \citep{Zhao+2016}, 
but with a $\zeta_0^{\rm H_2}$ close to the upper limit, only compact discs of 
$\lesssim$10~AU radius could form. Given the observational evidence of depletion 
of $\lesssim$10~nm VSGs in dense molecular cores \citep{Tibbs+2016}, and the 
typical range of CR ionization rate in dense cores (few 10$^{-18}$~s$^{-1}$ to 
few 10$^{-16}$~s$^{-1}$; \ct{Caselli+1998,Padovani+2009}), the new 
requirements on microphysics basically imply a universality of RSDs around 
low-mass protostars. The spread in disc radius and morphology of nearby protostellar 
sources is likely a result of different cloud initial conditions, especially the 
environmental factors such as CR ionization rate \citep{Kuffmeier+2020}. 
For example, sources with small discs (e.g., B335) may indicate a higher CR 
ionization rate at the cloud scale \citep{Yen+2019,Yen+2020}, while regions with 
$\zeta_0^{\rm H_2}$ close to the canonical value of 10$^{-17}$~s$^{-1}$ allow 
both small and large discs or multiple systems to form 
\citep{Tobin+2016b,Segura-Cox+2018}, depending on other initial parameters 
\citep{Zhao+2018a} such as magnetic field strength, rotation speed, 
density perturbation, as well as the degree of misalignment between the magnetic 
field and rotation. 

Finally, the recent discovery of a $\sim$10$^4$~AU streamer in the protostellar envelope 
of IRAS 03292+3039 \citep{Pineda+2020} opens a new window for studying protostellar 
collapse and disc formation. Since individual star-forming cores are often not isolated 
from the large-scale filaments and molecular clouds, it would be more self-consistent 
to investigate how such asymmetric accretion flows from larger scales could affect 
disc formation \citep{Kuffmeier+2017,Kuznetsova+2020} under the regulation of 
non-ideal MHD effects.

\subsection{Numerical Limitations}

Despite the d$t$ floors we imposed for both AD and Hall effect, such floors are 
small enough and only trigger inside the disc or within the innermost <10~AU 
(e.g., Fig.~\ref{Fig:2.4opt3eta}), where Ohmic dissipation already becomes 
very efficient so that the magnetic field strength saturates and the tendency of 
field bending by AD or Hall effect is suppressed (see detailed discussions in 
\PaperI). As demonstrated in \citet{Zhao+2018a}, the process of disc formation is 
governed by the efficiency of magnetic diffusion and angular momentum transport 
outside (instead of inside) the disc, the d$t$ floors used in the current study 
should not affect whether disc forms or not. Future zoom-in studies fully resolving 
the protoplanetary disc itself and magnetic diffusion therein are necessary to 
accurately follow the long-term disc evolution after the depletion of protostellar 
envelopes.

\section{Summary}
\label{Chap.Summary}

We have extended the study of Hall regulated disc formation in \PaperI~to 
a general case of disc formation enabled by non-ideal MHD effects. 
In particular, we have focused on the interplay between AD and Hall effect 
in the collapsing envelope, and the dependence on the magnetic field polarity 
and microphysics. We have found that, in spite of the non-trivial behaviors of 
the ambipolar and Hall drift, the combined effort of AD and Hall effect in the 
radial direction is to move the magnetic field radially outward relative to the 
infalling matter in the envelope, which greatly promotes the formation and survival 
of RSDs. Further conclusions are listed below. 
\begin{description}
\item 1. We confirm the early result of \citet{Li+2011} that disc formation is 
suppressed with the standard MRN size distribution ($a_{\rm min}=0.005~\mu$m, 
$a_{\rm min}=0.25~\mu$m), which results in a low Hall diffusivity within 
$\lesssim$100~AU scale and a low ambipolar diffusivity at 10$^2$--10$^3$~AU scale. 
Both the radial AD or the azimuthal Hall drift of magnetic fields are 
insufficient to enable disc formation. 
\item 2. Truncating the MRN size distribution at $a_{\rm min} \approx 0.03~\mu$m 
maximizes the Hall diffusivity in the inner envelope ($\lesssim$100--200~AU) and 
leads to a Hall dominated collapse. RSDs of $\sim$20--30~AU radii form regardless 
of the polarity of the magnetic field. However, the direction of disc rotation 
reverses as the magnetic field polarity flips: normally-rotating for anti-aligned 
cases ($\bmath{\Omega \cdot B}<0$) versus counter-rotating for aligned cases 
($\bmath{\Omega \cdot B}>0$), with respect to the direction of initial core rotation. 
\item 3. For the Hall dominated collapse, when $\bmath{\Omega \cdot B}>0$, the radial 
components of both Hall and ambipolar drift are pointing towards +$r$, cooperatively 
diffusing the magnetic field radially outward. When $\bmath{\Omega \cdot B}<0$, the 
radial Hall drift points towards +$r$ at 100~AU scale but reverses to -$r$ at tens of 
AU scale; however, the radial ambipolar drift along +$r$ always ensures the combined 
radial drift of the magnetic field relative to the bulk neutral matter is directed 
radially outward. 
\item 4. Further truncating the MRN size distribution at $a_{\rm min} \approx 0.1~\mu$m 
maximizes the ambipolar diffusivity throughout the collapsing envelope and leads to 
an AD dominated collapse. Disc sizes and morphologies are largely unaffected by Hall 
effect in this case and hence independent of the magnetic field polarity. The radial 
drift of magnetic fields is primarily determined by AD, i.e., pointing radially outward; 
the azimuthal Hall drift is weakened due to the reduced radial pinching of poloidal 
magnetic fields by AD. Efficient AD of magnetic fields often results in self-gravitating 
discs that would become large spiral structures or multiple systems in a full 3D set-up.
\item 5. Counter-rotating envelope only develops in the Hall dominated collapse, 
where Hall effect, by regulating the topology of the magnetic field, redistributes 
angular momentum among different parts of the envelope. Either a ``butterfly-shaped'' 
thin shell is counter-rotating (for $\bmath{\Omega \cdot B}<0$), or the inner envelope 
enclosing the disc and outflow region is counter-rotating (for $\bmath{\Omega \cdot B}>0$). 
The aligned case also shows a ``butterfly-shaped'' thin shell with an excess of 
angular momentum. 
\item 6. With the help of Hall effect, the requirements on microphysics for disc 
formation are somewhat relaxed in comparison to studies focused only on AD. 
The lower limit of grain size $a_{\rm min}$ reduces from a few times 10~nm to 
just $\lesssim$10~nm; the upper limit of $\zeta_0^{\rm H_2}$ at the core scale 
increases from a few 10$^{-17}$~s$^{-1}$ to 2--3$\times$10$^{-16}$~s$^{-1}$. 
Under such criteria, the majority of low-mass star-forming cores would allow RSDs to 
form. However, only compact discs of $\lesssim$10~AU radius could form if the parent 
core has a relatively high CR ionization rate 
($\zeta_0^{\rm H_2} \gtrsim 10^{-16}$~s$^{-1}$). 
\item 7. The polarity of the magnetic field can also affect the disc to stellar mass 
ratio and hence the disc stability in the Hall dominated collapse; the aligned case 
tends to grow a large stellar mass during its initial disc suppression phase. 
\item 8. In the aligned case, outflow cavity is generally wider and outflow speed 
is a few times faster than in the anti-aligned case, due to the strengthening of 
$B_\phi$ by the azimuthal Hall drift. 
The Hall-dominated collapse ($a_{\rm min} \approx 0.03~\mu$m) also launches stronger 
outflows than the AD-dominated collapse ($a_{\rm min} \approx 0.1~\mu$m), as poloidal 
magnetic fields are weaker and less pinched when AD is strong.
\item 9. Hall effect is only efficient in spining up/down the gas rotation in 
relatively strongly magnetized cores. For $\lambda \gtrsim 10$, the azimuthal Hall 
drift is negligible even when adopting $a_{\rm min} \approx 0.03~\mu$m.
\item 10. In general, if ambipolar and Hall drift are cooperative in a given direction 
($r$- or $\phi$- direction), they are counteractive in the orthognal direction ($\phi$- 
or $r$- direction). Such a principle can be applied to either a collapsing envelope or 
a rotating disc, with magnetic field lines being preferentially pinched radially by 
collapse in the former while bended azimuthally by rotation in the latter. 
The combined effect of AD and Hall effect is to drift the magnetic field against 
the primary direction of field bending.
\item 11. The sign of $\bmath{\Omega \cdot B}$ of protoplanetary discs are closely 
related to the microphysics in the protostellar envelope; only the anti-aligned 
configuration is possible for discs formed from a Hall dominated collapse 
whereas both aligned and anti-aligned configurations are allowed for discs fromed 
from an AD dominated collapse.
\end{description}

We conclude that the strong dependence of non-ideal MHD effects on microphysics 
places microphysics in a pivotal role in protostellar collapse and disc formation. 
The depletion of VSGs below $\lesssim$10~nm in prestellar cores remains critical 
for promoting disc formation. Slight changes in microphysics can leave profound 
imprints on observables, including rotation direction of envelopes and discs, 
size and morphology of discs, velocity and open-angle of outflows, and possibly 
disc to stellar mass ratio. Although most of these detailed features are related to 
Hall effect and the magnetic field polarity, AD remains as the cornerstone for 
disc formation and survival by ensuring a radially outward diffusion of the 
magnetic field in the collapsing envelope.

\section*{Acknowledgements}

BZ and PC acknowledge support from the European Research Council 
(ERC; project PALs 320620) and the Max-Planck Society. 
Z.-Y. L. and K.-H. L. are supported in part by NASA 80NSSC18K1095, 80NSSC20K0533, 
NSF 1716259. 
HS and RK acknowledge grant support from the ASIAA and the Ministry of Science 
and Technology in Taiwan through MOST 105-2119-M-001-037- and 105-2119-M-001-044-MY3. 
ZeusTW is developed and maintained by the CHARMS group in ASIAA. 
Numerical simulations are carried out on the CAS group cluster at MPE.







\bsp	
\label{lastpage}
\end{document}